\documentclass[preprint]{elsarticle}

\usepackage{lineno,hyperref}
\modulolinenumbers[5]

\journal{Journal of Parallel and Distributed Computing}

\usepackage{color}
\usepackage{amsmath}
\usepackage{array,multirow}
\usepackage{graphicx}
\usepackage{caption}
\usepackage{subcaption}
\usepackage{color}
\usepackage{mathtools}
\usepackage[inline]{enumitem}
\usepackage{threeparttable}
\usepackage{nicefrac}

\DeclarePairedDelimiter\floor{\lfloor}{\rfloor}
\DeclarePairedDelimiter\abs{\lvert}{\rvert}%
\DeclarePairedDelimiter\norm{\lVert}{\rVert}%

\newcommand{\rom}[1]{\uppercase\expandafter{\romannumeral #1\relax}}

\usepackage{bm}

\usepackage{booktabs}

\makeatletter
\let\oldabs\abs%
\def\abs{\@ifstar{\oldabs}{\oldabs*}}
\let\oldnorm\norm%
\def\norm{\@ifstar{\oldnorm}{\oldnorm*}}
\makeatother

\begin{document}

\begin{frontmatter}

\title{FFT, FMM, and Multigrid on the Road to Exascale: performance challenges and opportunities}

\author[UIUC]{Huda Ibeid\corref{corresponding}}
\cortext[corresponding]{Corresponding author}
\ead{hibeid@illinois.edu}

\author[UIUC]{Luke Olson}
\ead{lukeo@illinois.edu}

\author[UIUC]{William Gropp}
\ead{wgropp@illinois.edu}

\address[UIUC]{University of Illinois at Urbana-Champaign, Urbana, IL 61801}

\begin{abstract}
FFT, FMM, and multigrid methods are widely used fast and highly scalable solvers for elliptic PDEs. However, emerging large-scale computing systems are introducing challenges in comparison to current petascale computers. Recent efforts~\cite{Dongarra2011} have identified several constraints in the design of exascale software that include massive concurrency, resilience management, exploiting the high performance of heterogeneous systems, energy efficiency, and utilizing the deeper and more complex memory hierarchy expected at exascale. In this paper, we perform a model-based comparison of the FFT, FMM, and multigrid methods in the context of these projected constraints. In addition we use performance models to offer predictions about the expected performance on upcoming exascale system configurations based on current technology trends.
\end{abstract}


\end{frontmatter}



\section{Introduction}

Elliptic PDEs arise in many applications in computational science and engineering. Classic examples are found in computational astrophysics, fluid dynamics, molecular dynamics, plasma physics, and many other areas. The rapid solution of elliptic PDEs remains of wide interest and often represents a significant portion of simulation time.

The fast Fourier transform (FFT), the fast multipole method (FMM), and multigrid methods (MG) are widely used fast and highly scalable solvers for elliptic PDEs. The FFT, FMM, and MG methods have been used in a wide variety of scientific computing applications such as particle-in-cell methods, the calculation of long-range (electrostatic) interactions in many-particle systems, such as molecular dynamics and Monte Carlo sampling~\cite{Arnold2013}, and in signal analysis. The performance expectations of these methods helps guide algorithmic changes and optimizations to enable migration to exascale systems, as well as to help identify potential bottlenecks in exascale architectures. In addition, modeling helps assess the trade-offs at extreme scales, which can assist in choosing optimal methods and parameters for a given application and specific machine architecture.

Each method has advantages and disadvantages, and all have their place as PDE solvers. Generally, the FFT is used for uniform discretizations, FMM and geometric MG are efficient solvers on irregular grids with local features or discontinuities, and algebraic MG can handle arbitrary geometries, variable coefficients, and general boundary conditions.  The focus of this study is on FFT, FMM, and \textit{geometric} MG, although several observations extend to an algebraic setting as well~\cite{Bienz2016}.

One aim of the International Exascale Software Project (IESP) is to enable the development of applications that exploit the full performance of exascale computing platforms~\cite{Dongarra2011}. Although these exascale platforms are not yet fully specified, it is widely believed that they will require significant changes in computing hardware architecture relative to the current petascale systems. The IESP roadmap reports that technology trends impose severe constraints on the design of an exascale software. Issues that are expected to affect system software and applications at exascale are summarized as

\begin{description}

 \item[Concurrency:] Future supercomputing performance will depend mainly on increases in system scale. Processor counts of one million or more for current systems~\cite{top500} whereas exascale systems are likely to incorporate one billion processing cores, assuming GHz technology.  As a result, this $1000 \times$ increase in concurrency necessitates new paradigms for computing for large-scale scientific applications to ensure extrapolated scalability.

 \item[Resiliency:] The exponential increase in core counts expected at exascale will lead to increases in the number of routers, switches, interconnects, and memory systems. Consequently, resilience will be a challenge for HPC applications on future exascale systems.

 \item[Heterogeneity:] As accelerators advance in both performance and energy efficiency, heterogeneity has become a critical ingredient in the pursuit of exascale computing. Exploiting the performance of these heterogeneous systems is a challenge for many methods.

 \item[Energy:] Power is a major challenge. Current petascale systems would reach the level of 100 MW if extended to exascale\footnote{For example, the Piz Daint supercomputer, which is ranked third and tenth on the TOP500 and Green500 lists, respectively, has power efficiency of 10.398 GFLOPs/W. An exascale machine with the same power efficiency will require 96 MW per exaFLOP.}. This imposes design constraints on both the hardware and software to improve the overall efficiency. Likewise, exascale algorithms need to focus on maximizing the achieved ratio of performance to power/energy consumption (power/energy efficiency), rather than focusing on raw performance alone.

 \item[Memory:] The memory hierarchy is expected to change at exascale based on both new packaging capabilities and new technologies to provide the memory bandwidth and capacity required at exascale.  Changes in the memory hierarchy will affect programming models and optimizations, and ultimately performance.

\end{description}

In this manuscript, we perform model-based comparison of the FFT, FMM, and MG methods  vis-\`a-vis these challenges. We also use performance models to estimate the performance on hypothetical future systems based on current technology trends.
The rest of the manuscript is organized as follows. A short description of the FFT, FMM, and MG methods is provided in Section~\ref{sec:methods}. In the following sections, we present, compare, and discuss the performance of these methods relative to the exascale constraints imposed by technology trends. These constraints are: concurrency (Section~\ref{sec:concurrency}), resiliency (Section~\ref{sec:resiliency}), heterogeneity (Section~\ref{sec:heterogeneity}), energy (Section~\ref{sec:energy}), and memory (Section~\ref{sec:memory}). Observations and conclusions are drawn in Sections~\ref{sec:observations} and~\ref{sec:conclusions}, respectively.

\section{Methods}~\label{sec:methods}

In this section we provide a brief description of FFT, FMM, and MG, in order to establish notation and as preamble to the performance analysis.

\subsection{Fast Fourier Transform}

The FFT is an algorithm for computing the $N$-point Discrete Fourier Transform (DFT) with $\mathcal{O}(N \log N)$ computational complexity. Let $\bm{x}=(x_1,x_2,\dotsc,x_N)$ be a vector of $N$ complex numbers, the 1-D DFT of $\bm{x}$ is defined as
\begin{equation}
\hat{x}_k = \sum^N_{j=1} x_j e^{-i \frac{2\pi k}{N}j}.
\end{equation}
The 3-D FFT is performed as three successive sets of independent 1-D FFTs.

\subsubsection{Parallel Domain Decomposition}

To compute the parallel 3-D FFT, the computational domain is decomposed across processors. There are two popular decomposition strategies for parallel computation: the slab decomposition (1-D decomposition) and the pencil decomposition (2-D decomposition).

In the case of a slab decomposition a 3-D array is partitioned into slabs along one axis so that each processor consists of $\frac{\sqrt[3]{N}}{P} \times \sqrt[3]{N} \times \sqrt[3]{N}$ points. This decomposition scheme is unsuitable for massively parallel supercomputer as the number of processors that can be used is limited by the number of slabs.
In contrast, in pencil decomposition (a 2-D decomposition) a 3-D array is partitioned in two dimensions, which allows the number of processors to increase. Two of the three dimensions of the cube are divided by $\sqrt{P}$. Hence, each processor has $\frac{\sqrt[3]{N}}{\sqrt{P}} \times \frac{\sqrt[3]{N}}{\sqrt{P}} \times \sqrt[3]{N}$ points. A pencil decomposition is used in the current analysis.

\subsubsection{FFT Calculation Flow}

\begin{figure}
\centering
\includegraphics[width=0.8\textwidth]{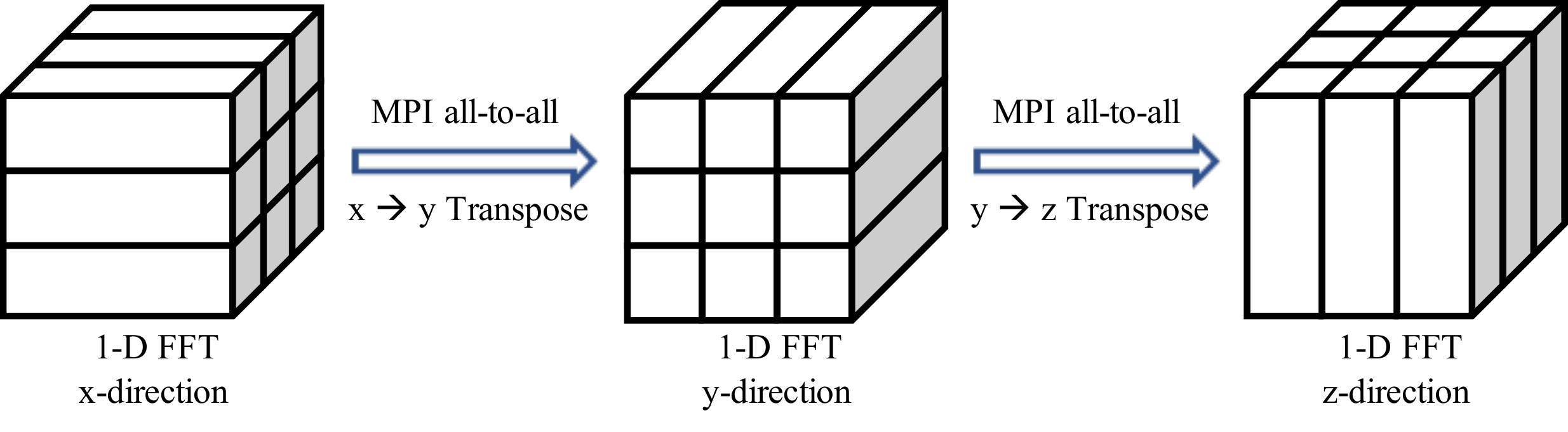}
\caption{Illustration of the 3-D FFT calculation flow using pencil decomposition.}\label{fig:fft_flow}
\end{figure}
The pencil decomposition of a 3-D FFT consists of three computation phases separated by two all-to-all communication phases. Each computation phase computes $\sqrt[3]{N} \times \sqrt[3]{N}$ 1-D FFTs of size $\sqrt[3]{N}$ in parallel. Each all-to-all communication requires $\mathcal{O}(\sqrt{P})$ exchanges for the transpose between pencil-shaped subdomains on $P$ processes. This calculation flow is illustrated in Figure~\ref{fig:fft_flow}.

The solution of the Poisson equation $-\Delta u = f$ based on FFT is
\begin{equation}
x = \hat{x}^{-1}(\hat{f}/ \abs{k}^2),
\end{equation}
where $\hat{f}$ is the Fourier transform of $f$ and $\hat{x}^{-1}$ is the inverse Fourier transform of $x$. Thus, solving the Poisson equation using Fourier transform can be broken down into three steps:
\begin{enumerate*}[label={\arabic*)}]
\item compute the FFT of $f$;
\item scale $\hat{f}$ by $\abs{k}^2$ in Fourier space; and
\item compute the inverse Fourier transform of the result.
\end{enumerate*}

\subsection{Fast Multipole Method}\label{subsec:fmm}

$N$-body problems are used to simulate physical systems of particle interactions under a physical or electromagnetic field~\cite{greengard1987}. The $N$-body problem can be represented by the sum
\begin{equation}
f(y_j) = \sum_{i=1}^N w_i K(y_j,x_i),
\end{equation}
where $f(y_j)$ represents a field value evaluated at a point $y_j$ that is generated by the influence of sources located at the set of centers $\{x_i\}$. $\{x_i\}$ is the set of source points with weights given by $w_i$, $\{y_j\}$ is the set of evaluation points, and $K(y,x)$ is the kernel that governs the interactions between evaluation and source points.

The direct approach to simulate the $N$-body problem evaluates all pair-wise interactions among the points which results in a computational complexity of $\mathcal{O}(N^2)$. This complexity is prohibitively expensive even for modestly large data sets. For simulations with large data sets, many faster algorithms have been invented, e.g., tree code~\cite{Barnes1986} and fast multipole methods~\cite{greengard1987}. The fast algorithms cluster points at successive levels of spatial refinement. The tree code clusters the far points and achieves $\mathcal{O}(N \log N)$ complexity. The further apart the points, the larger the interaction groups into which they are clustered. On the other hand, FMM divides the computational domain into near-domain and far-domain and computes interactions between clusters by means of local and multipole expansions, providing $\mathcal{O}(N)$ complexity.  Other $N$-body approaches follow a similar strategy~\cite{Beams2016,Burstedde2011}. FMM is more than an $N$-body solver, however. Recent efforts to view the FMM as an elliptic PDE solver have opened the possibility to use it as a preconditioner for even a broader range of applications~\cite{Ibeid2018}.

\subsubsection{Hierarchical Domain Decomposition}

\begin{figure}
\centering
	\begin{subfigure}[t]{0.2\textwidth}
        \centering
        \includegraphics[width=\textwidth]{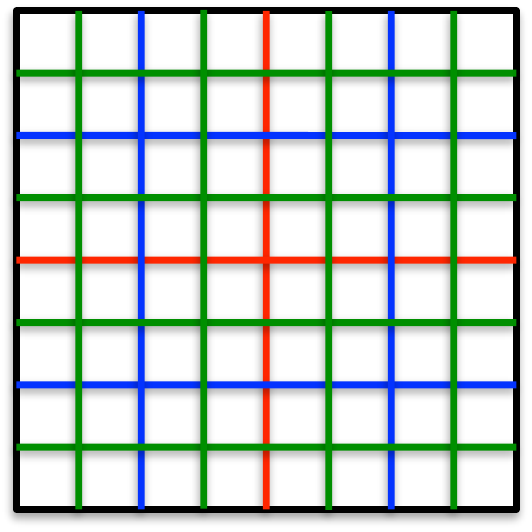}
        \caption{2-D domain}\label{fig:domain}
    \end{subfigure}
	\begin{subfigure}[t]{0.75\textwidth}
        \centering
        \includegraphics[width=\textwidth]{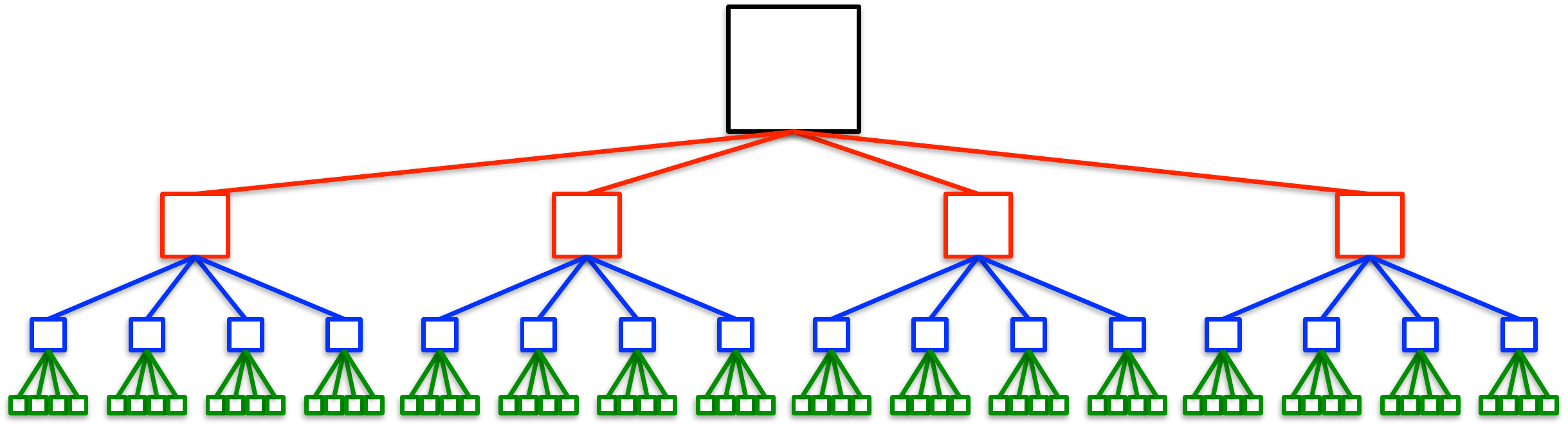}
        \caption{Quad-tree}\label{fig:tree}
    \end{subfigure}
    \caption{Decomposition of a 2-D computational domain into a quad-tree.}\label{fig:decomposition}
\end{figure}
The first step of the FMM algorithm is the decomposition of the computational domain. This spatial decomposition is accomplished by a hierarchical subdivision of the space associated with a tree structure. The 3-D spatial domain of FMM is represented by oct-trees, where the space is recursively subdivided into eight boxes until the finest level of refinement or ``leaf level''.  Figure~\ref{fig:decomposition} illustrates an example of a hierarchical space decomposition for a 2-D domain that is associated with a quad-tree structure.

\subsubsection{The FMM Calculation Flow}

\begin{figure}
\centering
\includegraphics[width=0.8\textwidth]{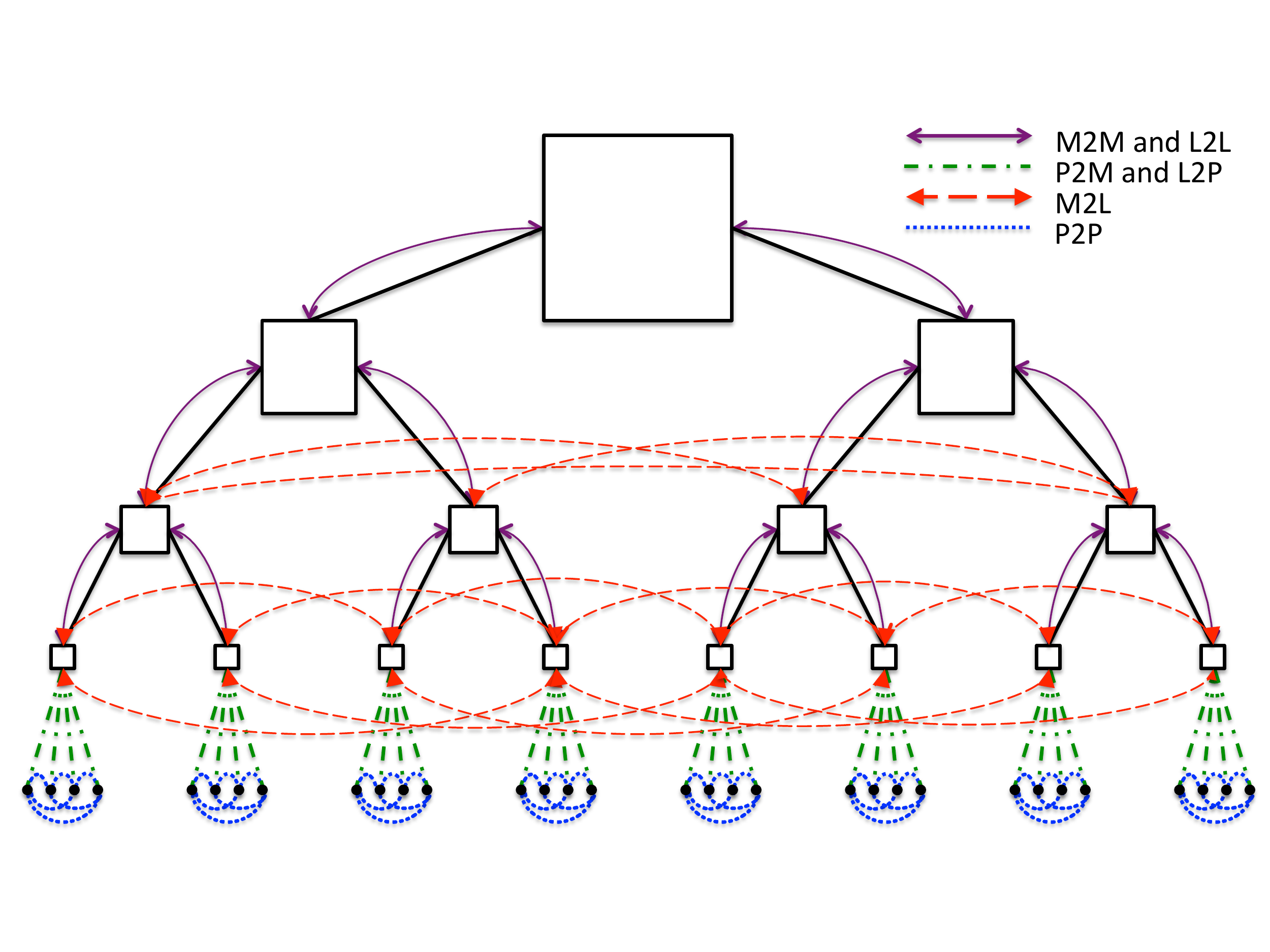}
\caption{Illustration of the FMM kernels: P2M (Point-to-Multipole), M2M (Multipole-to-Multipole), M2L (Multipole-to-Local), L2L (Local-to-Local), L2P (Local-to-Point), and P2P (Point-to-Point).}\label{fig:fmm_flow}
\end{figure}
The FMM calculation begins by transforming the mass/charge of the source points into multipole expansions by means of a Point-to-Multipole kernel (P2M). Then, the multipole expansions are translated to the center of larger boxes using a Multipole-to-Multipole kernel (M2M). FMM calculates the influence of the multipoles on the target points in three steps: (1) translation of the multipole expansions to local expansions between well-separated boxes using a Multipole-to-Local kernel (M2L); (2) translation of local expansions to smaller boxes using a Local-to-Local kernel (L2L); and (3) translation of the effect of local expansions in the far field onto target points using a Local-to-Point kernel (L2P). All-pairs interaction is used to calculate the near field influence on the target points by means of a Point-to-Point kernel (P2P). Figure~\ref{fig:fmm_flow} illustrates the FMM main kernels: Point-to-Multipole (P2M), Multipole-to-Multipole (M2M), Multipole-to-Local (M2L), Local-to-Local (L2L), Local-to-Point (L2P), and Point-to-Point (P2P). The dominant kernels of the FMM calculation are P2P and M2L.

\subsubsection{FMM Communication Scheme}

In this study, we adopt a tree structure that is similar to the one described in~\cite{Ibeid2016, Abduljabbar2017} where FMM uses a separate tree structure for the local and global trees. Each leaf of the global tree is a root of a local tree for a particular MPI process. Therefore, the depth of the global tree depends only on the number of processes $P$ and grows with $\log_8(P)$ in 3-D. Each MPI process stores only the local tree, which depth grows with $\log_8(N/P)$, and communicates the halo region at each level of the local and global tree. Table~\ref{t:FMMcomm} shows the number of boxes that are sent at the ``Global M2L'', ``Local M2L'', and ``Local P2P'' phases where $i$ refers to the level in the local tree and $26$ is the number of nearest neighbors.
\begin{table}[t]
\centering
\caption{Amount of communication in FMM.}\label{t:FMMcomm}
\small
\begin{tabular}{cc}
 \toprule
  & Boxes to send / level \\ [0.5ex]
 \midrule
Global M2L & $26\times8$ \\
Local M2L & ${(2^i+4)}^3-8^i$ \\
Local P2P & ${(2^i+2)}^3-8^i$ \\
\bottomrule
\end{tabular}
\end{table}

\subsection{Multigrid}

Multigrid methods are among the most effective solvers for a wide range of problems. They target the solution of a sparse linear system $A x = b $ with $N$ unknowns in a computational complexity of $\mathcal{O}(N)$. The basic idea behind MG is to use a sequence of coarse grids to accelerate convergence of the fine grid solution. The building blocks of the multigrid method are the smoothing, restriction, and interpolation operators. These are usually 3-D stencil operations on a structured grid in the case of geometric multigrid (GMG) and sparse matrix-vector multiplications (SpMV) in algebraic multigrid (AMG). In the current study, multigrid refers to the geometric multigrid.

\begin{figure}
\centering
\includegraphics[width=0.8\textwidth]{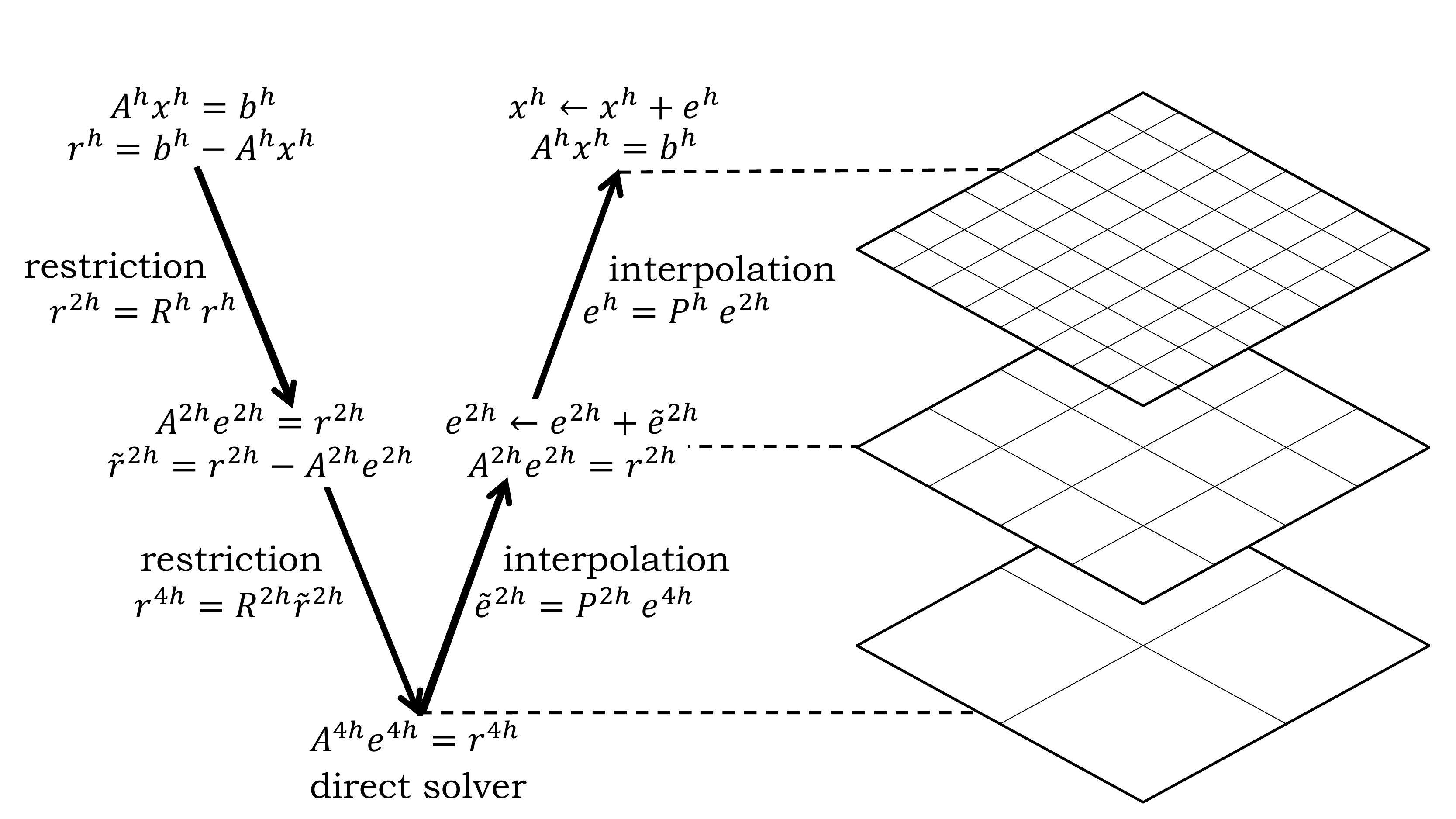}
\caption{Illustration of the multigrid V-cycle.}\label{fig:mg_flow}
\end{figure}
The V-cycle, shown in Figure~\ref{fig:mg_flow}, is the standard process of a multigrid solver.  Starting at the finest structured grid, a smoothing operation is applied to reduce high-frequency errors followed by a transfer of the residual to the next coarser grid. This process is repeated until the coarsest level is reached, at which point the linear system is solved with a direct solver. The error is then interpolated back to the finest grid. The V-cycle is mainly dominated by the smoothing and residual operations on each level.

A multigrid solver is constructed by repeated application of a V-cycle. The number of V-cycles required to reduce the norm of the error by a given tolerance $\epsilon$ is estimated by
\begin{equation}
itr_{\!_{\textnormal{MG}}} = \frac{\log \epsilon}{\log \rho},
\end{equation}
where $\rho$ is the convergence rate. Generally, the convergence rate is bounded by ${((\kappa - 1)/\kappa)}^\mu$ where $\mu$ is the number of smoothing steps and $\kappa$ is the condition number of the matrix $A$~\cite{Bank1985}.

\section{Exascale Projection}

In this paper we consider exascale systems built from hypothetical processors based on extrapolating current technology trends. This section describes how we project these hypothetical CPU-based and GPU-based exascale systems. Similar concept was applied in 2010 by~\cite{Czechowski2012}.

\begin{figure}
\centering
	\begin{subfigure}[t]{0.45\textwidth}
        \centering
        \includegraphics[width=\textwidth]{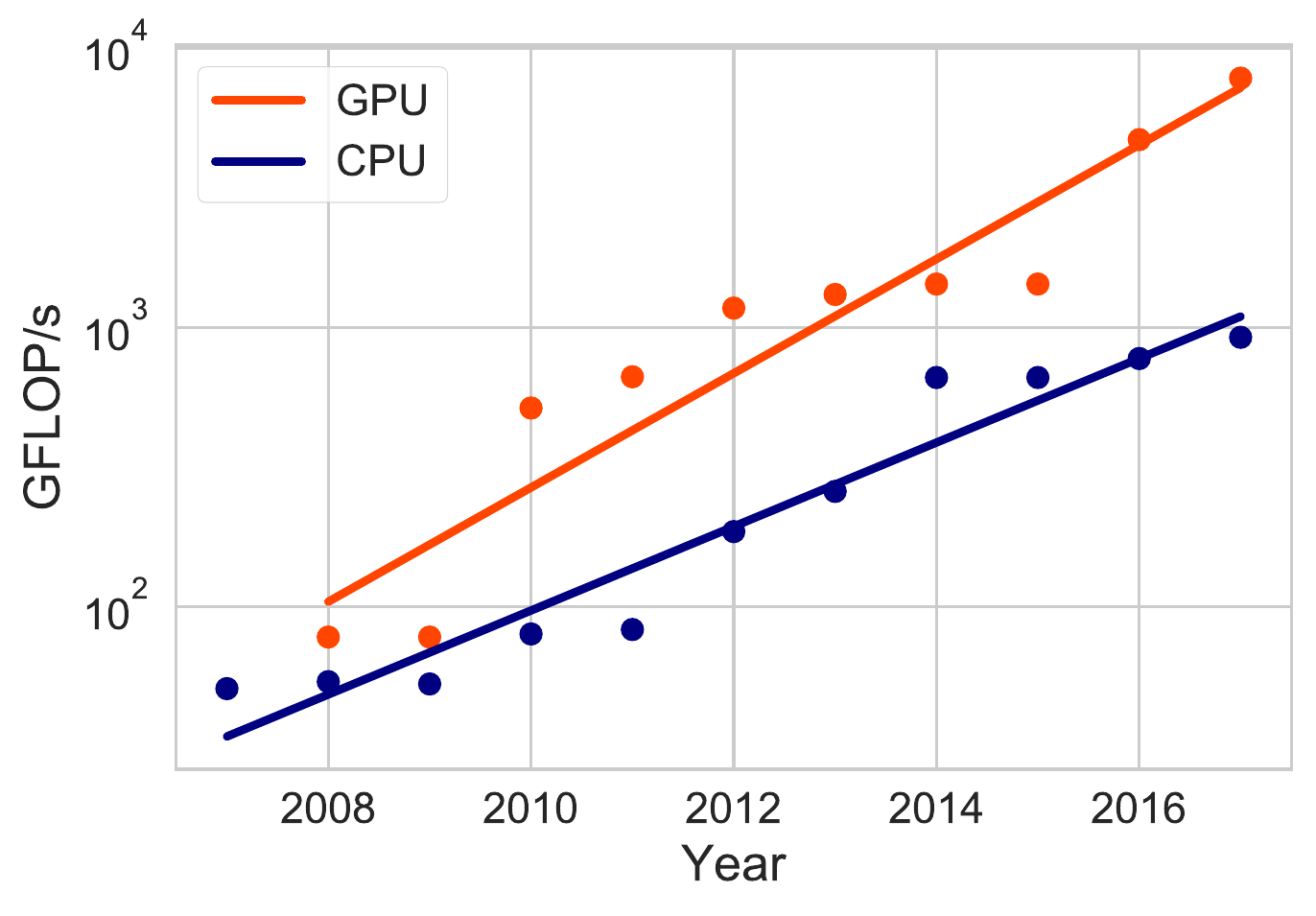}
        \caption{Processor performance}
    \end{subfigure}
	\begin{subfigure}[t]{0.45\textwidth}
        \centering
        \includegraphics[width=\textwidth]{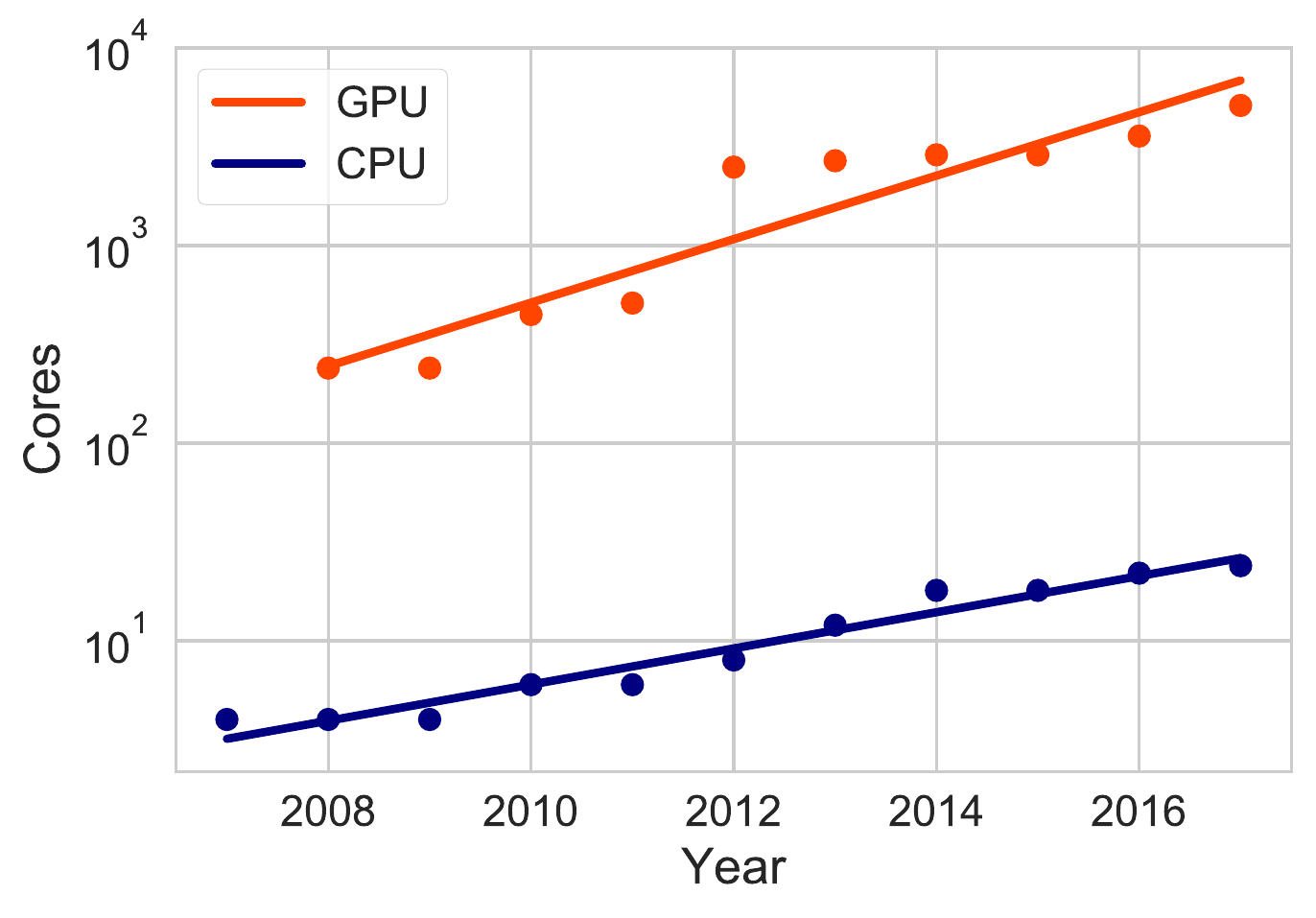}
        \caption{Cores per processor}
    \end{subfigure}
	\\
	\begin{subfigure}[t]{0.45\textwidth}
        \centering
        \includegraphics[width=\textwidth]{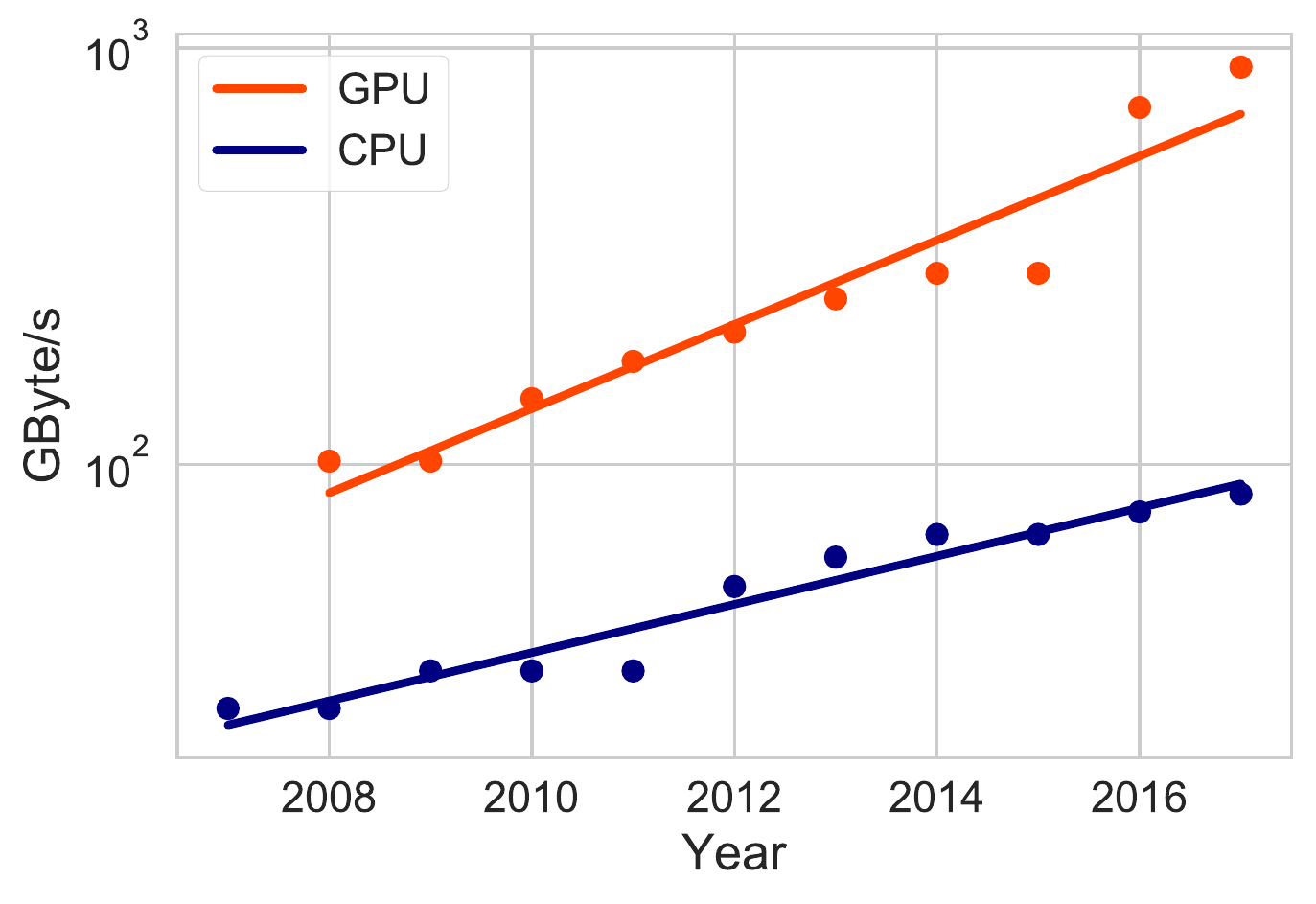}
        \caption{Memory bandwidth}
    \end{subfigure}
	\begin{subfigure}[t]{0.45\textwidth}
        \centering
        \includegraphics[width=\textwidth]{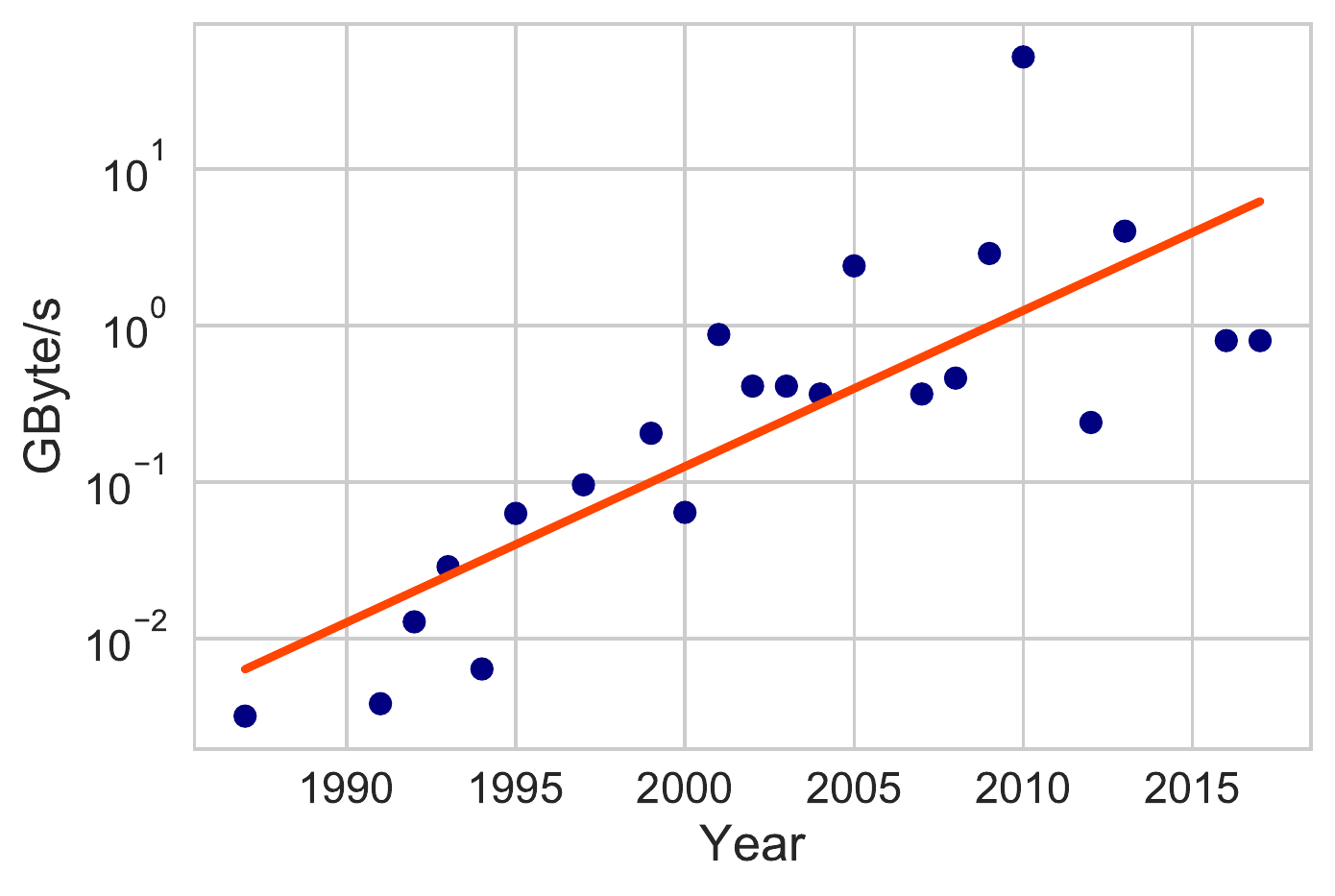}
        \caption{Link bandwidth}
    \end{subfigure}
    \caption{Hardware trends.}\label{fig:trends}
\end{figure}
We collect CPUs and GPUs peak performance, memory bandwidth, and  number of cores per processor for the period $2007-2017$. Linear regression is then used to find the doubling-time estimate for each parameter, as shown in Figure~\ref{fig:trends}. For the network link bandwidth, we begin with the data collected in~\cite{Czechowski2012}, which covers the period $1986-2012$. We then collect the same data for systems that made the TOP500 list since 2012.

\begin{figure}
\centering
\includegraphics[width=0.6\textwidth]{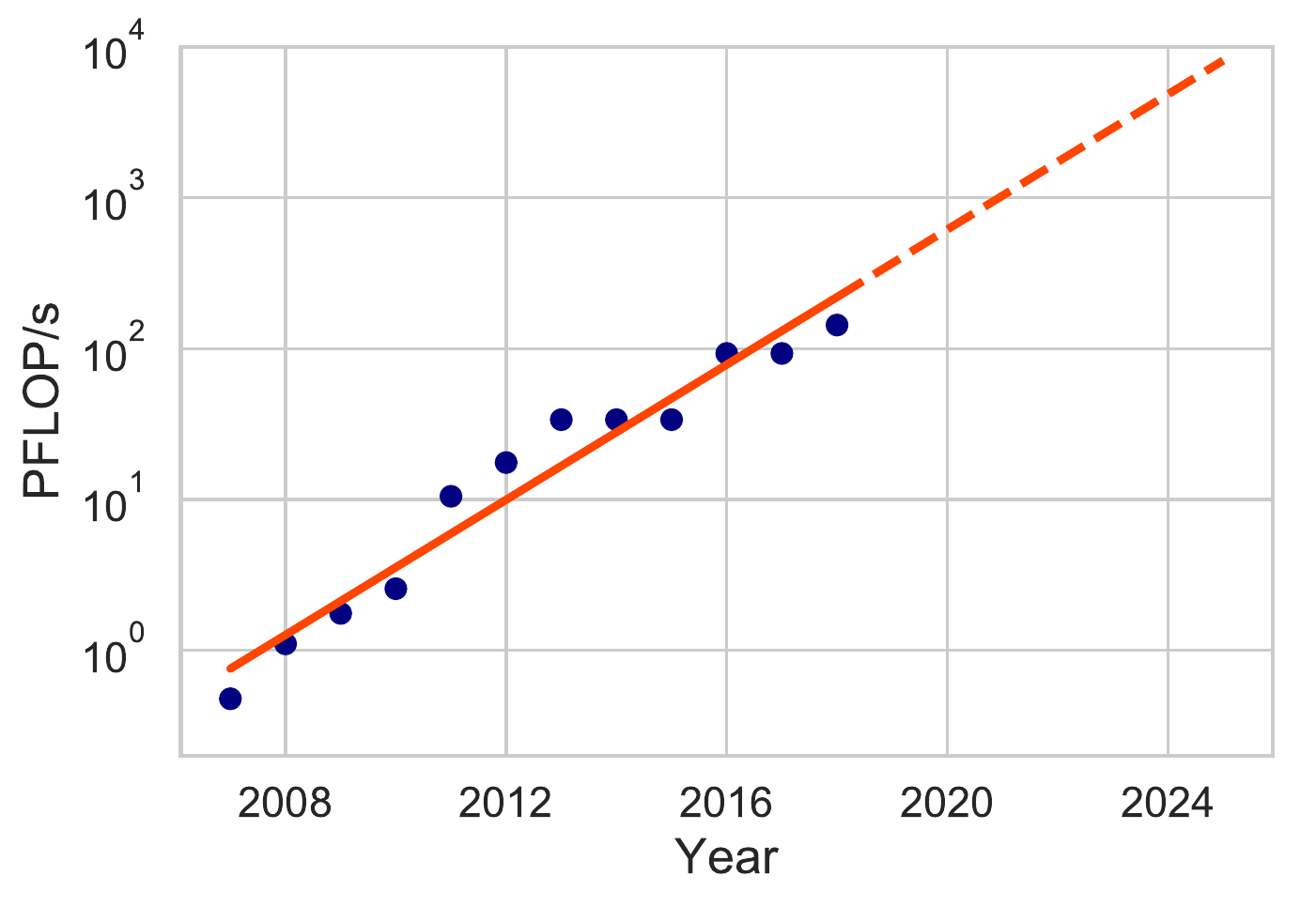}
\caption{Top machine on the TOP500 list by year~\cite{top500}.}\label{fig:top500}
\end{figure}
Table~\ref{t:exa} shows processor architecture projections, from starting values on the Argonne National Laboratory's Cooley (GPU-based) and KAUST’s Shaheen~\rom{2} system (CPU-based), both delivered in 2015. We assume that in 2025, we will be able to build a 7 exaFLOP/s (double-precision) system, as projected in Figure~\ref{fig:top500}. The value ``Processors'' count in Table~\ref{t:exa} is scaled to reflect this performance.

\begin{table}
\centering
\begin{threeparttable}
\caption{Processor architecture projections.}
\label{t:exa}
\small
\begin{tabular}{cccccc}
 \toprule
\multicolumn{2}{c}{\multirow{2}{*}{Parameter}} & 2015 values & Doubling time & 10-year\tnote{1} & \multirow{2}{*}{Value}\\ [0.5ex]
\multicolumn{2}{c}{} & & (in years) & increase factor & \\ [0.5ex]
\midrule
Processor & $1/t_{\textnormal{cpu}}$ & 588.8 GF/s & 2.0 & 32$\times$ & 18.8 TF/s \\
peak & $1/t_{\textnormal{gpu}}$ & 1.45 TF/s & 1.47 & 111.6$\times$ & 161.8 TF/s \\
\midrule
Memory & $1/\beta_{\textnormal{cpu}}$ &  68 GB/s & 5.2 & 3.8$\times$ & 258 GB/s \\
bandwidth & $1/\beta_{\textnormal{gpu}}$ & 240 GB/s & 2.98 & 10.2$\times$ & 2.4 TB/s \\
\midrule
\multirow{2}{*}{Cores} & $\rho _{\textnormal{cpu}}$ &16 & 3.29 & 8.2$\times$ & 132 \\
& $\rho_{\textnormal{gpu}}$ & 2,496 & 1.87 & 40.7$\times$ & 101.6k \\
\midrule
Fast & $Z_{\textnormal{cpu}}$ & 40 MB & \multirow{2}{*}{2.0} & \multirow{2}{*}{32.0$\times$} & 1.3 GB \\
memory & $Z_{\textnormal{gpu}}$ & 1.5 MB & & & 48 MB \\
\midrule
\multirow{2}{*}{Line size} & $L_{\textnormal{\textnormal{cpu}}}$ & 64 B & \multirow{2}{*}{10.2} & \multirow{2}{*}{2.0$\times$} & 128 B\\
& $L_{\textnormal{gpu}}$ & 128 B & & & 256 B\\
\midrule
Link & \multirow{2}{*}{$1/\beta_{\textnormal{link}}$} & \multirow{2}{*}{10 GB/s} & \multirow{2}{*}{3.0} & \multirow{2}{*}{10$\times$} & \multirow{2}{*}{100 GB/s} \\
bandwidth & & & & &\\
\bottomrule
Machine & \multirow{2}{*}{$R_{\textnormal{peak}}$} & \multirow{2}{*}{7 PF/s} & \multirow{2}{*}{1.0} & \multirow{2}{*}{1000.0$\times$} & \multirow{2}{*}{7 EF/s} \\
peak & & & & & \\
\midrule
Processors & $P _{\textnormal{cpu}}$ & 11,889  & 2.01 & 31.3$\times$ & 372k \\
($R _{\textnormal{peak}} \times t$) & $P_{\textnormal{gpu}}$ & 4,828 & 3.15 & 9$\times$ & 43.3k \\
\bottomrule
\end{tabular}
\begin{tablenotes}
\item[1] The 10-year increase factor is calculated using $2^{\nicefrac{10}{\textnormal{Doubling Time}}}$.
\end{tablenotes}
\end{threeparttable}
\end{table}

\section{Concurrency}\label{sec:concurrency}

To gain some insight into the solvers' performance on the massively concurrent systems expected at exascale, we derive analytical performance models that include computation and both intra- and inter-node communication costs. The intra-node communication along with the computation cost account for the single node performance which is a critical building-block in scalable parallel programs, whereas the inter-node term reflects the impact of network communication on the scalability.

In this section we develop performance models for FFT, FMM, and MG on $P$ nodes for a total problem size of $N = \sqrt[3]{N} \times \sqrt[3]{N} \times \sqrt[3]{N}$. Throughout, the computation time is defined as the total number of floating-point operations, multiplied by the time per floating-point operation, $t_c$, in seconds. Memory movement is modeled as the total data fetched into fast memory, multiplied by the memory bandwidth inverse ($\beta_{\textnormal{mem}}$) in units of seconds per element. Assuming arithmetic and memory operations are not overlapped, the total execution time $T_{\textnormal{exe}}$ is given by
\begin{equation}
T_{\textnormal{exe}}  =  T_{\textnormal{comp}} + T_{\textnormal{mem}},
\end{equation}
and with overlap, $T_{\textnormal{exe}}$ is given by
\begin{equation}
T_{\textnormal{exe}}  \approx \max(T_{\textnormal{comp}},T_{\textnormal{mem}}),
\label{eq:overlap}
\end{equation}
where $T_{\textnormal{comp}}$ is the computation time and $T_{\textnormal{mem}}$ is the time spent transferring data in a two-level memory hierarchy between the main memory and cache. $T_{\textnormal{exe}}$ in~\eqref{eq:overlap} can be rewritten as
\begin{equation}
T_{\textnormal{exe}} = n_{\textnormal{FLOP}} \cdot t_c \cdot \max(1, \frac{B_{\tau}}{AI}),
\label{eq:T}
\end{equation}
where $n_{\textnormal{FLOP}}$ is the number of FLOPs, $n_{\textnormal{mem}}$ is the number of main memory operations, $B_{\tau} \equiv \beta_{\textnormal{mem}} /t_c$ is the processor time balance, and $AI \equiv n_{\textnormal{FLOP}} / n_{\textnormal{mem}}$ is the arithmetic intensity. In order to minimize the execution time, $AI$ must be larger than $B_{\tau} $. This condition is referred to as the balance principle.

Inter-node communication cost is modeled using the postal model or $\alpha$--$\beta_{\textnormal{link}}$ model for communication, where $\alpha$ represents communication latency, $\beta_{\textnormal{link}}$ is the send time per element over the network (inverse the link bandwidth). Using this basic model, communication cost can be represented as
\begin{equation}
T_{\textnormal{net}}= m \alpha+n\beta_{\textnormal{link}},
\end{equation}
where $m$ and $n$ are the maximum number of messages and total number of elements sent by a process, respectively.

\subsection{Fast Fourier Transform}

\subsubsection{Computation Costs}

The 1-D Cooley-Tukey FFT of size $\sqrt[3]{N}$ consists of approximately $(5\sqrt[3]{N}\log \sqrt[3]{N})$ floating-point operations. Hence, the total computation time of the 3-D FFT is
\begin{equation}
  T_{\textnormal{comp},\textnormal{FFT}} = 3 \cdot \frac{5N \log \sqrt[3]{N}}{P} \cdot t_c,
\end{equation}
This model accounts for the three computational phases where each phase consists of $\sqrt[3]{N} \times \sqrt[3]{N}$ 1-D FFTs performed in parallel.

\subsubsection{Memory Access Costs}

For a cache with size $Z$ and cache-line length $L$ in elements, a cache-oblivious $\sqrt[3]{N}$-point 1-D FFT incurs $\Theta(1+\frac{\sqrt[3]{N}}{L}(1+\log_Z \sqrt[3]{N}))$ cache misses, for each transferring line of size $L$~\cite{Czechowski2012, Frigo2012}. This bound is optimal, matching the lower bound by Hong and Kung~\cite{Wei1981} when $\sqrt[3]{N}$ is an exact power of two. Thus, the time spent moving data between the main memory and a processor in the 3-D FFT is given by
\begin{equation}
T_{\textnormal{mem},\textnormal{FFT}} = 3 \cdot \frac{N \log_Z \sqrt[3]{N}}{P} \cdot \beta_{\textnormal{mem}}.
\end{equation}

\subsubsection{Network Communication Costs}

In the pencil decomposed 3-D FFT, each processor performs two all-to-all communications with $\sqrt{P}$ other processors sending a total of $\frac{N}{P}$ data points at each communication phase. Hence, the FFT inter-node communication time is approximated by
 \begin{equation}
 T_{\textnormal{net},\textnormal{FFT}} =  2 \cdot (\sqrt{P} \cdot \alpha + \frac{N}{P} \cdot \beta_{\textnormal{link}}),
 \end{equation}
where the factor of two accounts for the two communication phases. Since a fully connected network is unlikely at exascale, a more realistic estimation of the communication cost must include the topology of the interconnect~\cite{Czechowski2012}. For example, on a 3-D torus network without task-aware process placement, the communication time is bounded by the bisection bandwidth $\frac{P^{2/3}}{\beta_{\textnormal{link}}}$. Thus
\begin{equation}
T_{\textnormal{net},\textnormal{FFT}} =  2 \cdot (\sqrt{P} \cdot \alpha + \frac{N}{P^{2/3}} \cdot \beta_{\textnormal{link}}).
\end{equation}

\subsection{Fast Multipole Method}

In this section, we present analytical models for the two phases of FMM that consume most of the calculation time\@: P2P and M2L\@. We assume a nearly uniform points distribution and therefore a full oct-tree structure.

\subsubsection{Computation Costs}

\paragraph{\bf P2P}
Assuming $q$ points per leaf box, the computational complexity of the P2P phase is $27 q^2 \frac{N}{q}$ where $27$ is the number of neighbors including the box itself. This leads to a computation cost of
\begin{equation}
T_{\textnormal{comp},\textnormal{P2P}} = 27 \cdot \frac{qN}{P} \cdot t_c.
\end{equation}

\paragraph{\bf M2L} The asymptotic complexity of the M2L phase depends on the order of expansion $k$ and the choice of series expansion. Table~\ref{t:M2Lcomplexity} shows the asymptotic arithmetic complexity with respect to $k$ for different expansions used in fast $N$-body methods~\cite{Yokota2013a,Yokota2017}.
\begin{table}[t]
\centering
\caption{Asymptotic arithmetic complexity with respect to the order of expansion $k$ for the different series expansions (3-D).}\label{t:M2Lcomplexity}
\small
\begin{tabular}{ccc}
 \toprule
 Type of expansion & Complexity \\ [0.5ex]
 \midrule
Cartesian Taylor & $\mathcal{O}(k^6)$ \\
Cartesian Chebychev & $\mathcal{O}(k^6)$ \\
Spherical harmonics  & $\mathcal{O}(k^4)$ \\
Spherical harmonics+rotation & $\mathcal{O}(k^3)$ \\
Spherical harmonics+FFT & $\mathcal{O}(k^2 \log^2k)$ \\
Planewave & $\mathcal{O}(k^3)$ \\
Equivalent charges & $\mathcal{O}(k^4)$ \\
Equivalent charges+FFT  & $\mathcal{O}(k^3\log k)$ \\
 \bottomrule
\end{tabular}
\end{table}

The kernel-independent FMM (KIFMM)~\cite{Ying2004}, which uses equivalent charges and FFT, has a more precise operations count of $k^3\log k+189 k^3$~\cite{Chandramowlishwaran2012}. Hence, the M2L phase of the KIFMM has a total computation cost of
 \begin{equation}
 T_{\textnormal{comp},\textnormal{M2L}} = \frac{N k^3\log k}{q \cdot P} \cdot t_c + 189 \cdot \frac{N k^3}{q \cdot P} \cdot t_c,
 \end{equation}
where 189 is the number of well-separated neighbors per box ($6^3-3^3=189$).

Another state-of-the-art FMM implementation is exaFMM~\cite{Yokota2013a} which uses Cartesian series expansion. ExaFMM has operations count of $189 k^{6}$. Hence
 \begin{equation}
 T_{\textnormal{comp},\textnormal{M2L}} = 189 \cdot \frac{N k^6}{q \cdot P} \cdot t_c.
 \end{equation}

\subsubsection{Memory Access Costs}

As shown in~\cite{Chandramowlishwaran2012}, the outer loops of the P2P and M2L computations can be modeled as sparse matrix-vector multiplies. A cache-oblivious algorithm~\cite{Blelloch2010} for multiplying a sparse $H \times H$ matrix with $h$ non-zeros by a vector establishes an upper bound on cache misses in the SpMV as
\begin{equation}\label{eq:SpMV}
\mathcal{O} \left(\frac{h}{L}+\frac{H}{Z^{1/3}}\right),
\end{equation}
for each transferring line of size $L$.

\paragraph{\bf P2P}
Applying~\eqref{eq:SpMV} gives an upper bound on the number of cache misses for the P2P phase as follows
\begin{equation}\label{eq:P2P_mem}
Q_{\textnormal{P2P}} \leq 4 \cdot \frac{N}{L \cdot P}+ b_{\!_{\textnormal{P2P}}} \cdot \frac{N/q}{L \cdot P}+ 4 \cdot \frac{N}{L \cdot P}+\frac{N/q}{{(\frac{Z}{4q})}^{\!^{1/3}} \cdot P},
\end{equation}
where $b_{\!_{\textnormal{P2P}}}$ is the average number of source boxes in the neighbor list of a target leaf box ($b_{\!_{\textnormal{P2P}}} = 26$ for an interior box in a uniform distribution). The first two terms on the right-hand side of~\eqref{eq:P2P_mem} refer to read access for the source boxes and the neighbor lists for each target box, while the third term refers to the update access for the target leaf box potentials. In P2P communication, coordinates and values of every point belonging to the box must be sent, resulting in a multiplication factor of four. We model the dominant access time as
\begin{equation}
T_{\textnormal{mem},\textnormal{P2P}} = \frac{N}{P} \cdot \beta_{\textnormal{mem}} + \frac{NL}{(Z^{(1/3)}q^{(2/3)}) \cdot P} \cdot \beta_{\textnormal{mem}}.
\end{equation}

\paragraph{\bf M2L}
Applying~\eqref{eq:SpMV} for the M2L phase gives an upper bound on the number of cache misses as follows
\begin{equation}\label{eq:M2L_mem}
Q_{\textnormal{M2L}} \leq \frac{(b_t+bs) f(k)}{L}+\frac{b_{\!_{\textnormal{M2L}}}b_t}{L}+\frac{b_t}{{\big(\frac{\bar{Z}}{f(p)}\big)}^{\!^{1/3}}},
\end{equation}
where $b_t$ is the number of target boxes, $b_s$ is the number of source boxes, $b_{\!_{\textnormal{M2L}}}$ is the average number of source boxes in the well-separated list of a target box ($b_{\!_{\textnormal{M2L}}} = 189$ for an interior box in a uniform distribution), $f(k)$ is the asymptotic complexity given in Table~\ref{t:M2Lcomplexity}, and $\bar{Z}$ is the effective cache size. For $\bar{Z}$, we assume that all possible M2L translation operators ($7^3-3^3=316$) are computed and stored in the cache, leading to $\bar{Z} = Z - 316 \cdot f(k)$.

Considering the higher order terms, the memory access cost of the M2L phase can be approximated by
\begin{equation}
T_{\textnormal{mem},\textnormal{M2L}} = \frac{N f(k)}{q \cdot P} \cdot \beta_{\textnormal{mem}} + \frac{N {f(k)}^{1/3} L}{q \bar{Z}^{1/3} \cdot P} \cdot \beta_{\textnormal{mem}}.
\end{equation}

\subsubsection{Network Communication Costs}

\paragraph{\bf P2P}

The P2P communication is executing only at the lowest level of the FMM tree where each node communicates with its 26 neighbors. In total, each node communicates one layer of halo boxes which create a volume of ${(2^l+2)}^3-8^l$ where $l= \log_8(N/P)$. Using the $\alpha$--$\beta_{\textnormal{link}}$ model, the inter-node communication cost of the P2P phase can be represented by
 \begin{equation}
 T_{\textnormal{net},\textnormal{P2P}} = 26 \alpha + n_{\!_{\textnormal{P2P}}} \beta_{\textnormal{link}},
 \end{equation}
where $n_{\!_{\textnormal{P2P}}}$ is the number of elements in $(((\frac{N}{P})^\frac{1}{3} +2)^3-\frac{N}{P})$ leaf boxes.

\paragraph{\bf M2L} Similar to the P2P phase, the number of communicating nodes in the M2L phase is always the 26 neighbors. To use the $\alpha$--$\beta_{\textnormal{link}}$ model, we estimate the amount of data that is sent at each level of the FMM hierarchy. Table~\ref{t:FMMcomm} shows the number of boxes that are sent at the ``Global M2L'', ``Local M2L'', and ``Local P2P'' phases where $i$ refers to the level in the local tree. Thus, the communication cost of the M2L phase at level $l$ is represented by
 \begin{equation}
 T^l_{\textnormal{net},\textnormal{M2L}} = 26 \alpha + n^l_{\!_{\textnormal{M2L}}} \beta_{\textnormal{link}},
 \end{equation}
where $n^l_{\!_{\textnormal{M2L}}}$ is the number of elements sent at level $l$.

\subsection{Multigrid}

The basic building blocks of the classic geometric multigrid algorithm are all essentially stencil computations. In this section, the multigrid solve time is modeled as the sum of the time spent smoothing, restricting, and interpolating at each level as follows
\begin{equation}
T^{\textnormal{MG}}_{\textnormal{solve}} = T_{S} + T_{R} + T_{I},
\end{equation}
where $T_{S}$, $T_{R}$, and $T_{I}$ are the smoothing, restriction, and interpolation times, respectively.

In the classical multigrid there are more grid points than processors at fine levels. Hence, all processors are active. On coarse grids, however, there are fewer grid points than processors. Therefore, some processors execute on one grid point while others are idle. Some approaches to alleviate this problem include redistributing the coarsest problem to a single process and redundant data distributions. We assume a na\"ive multigrid implementation where the number of points decreases by a constant factor $\gamma$ in each dimension after each restriction operation. For simplicity of analysis, we assume restriction and interpolation require only communication with neighbors~\cite{Gahvari2010}.

\subsubsection{Computation Costs}

The smoother is a repeated stencil application. Each smoothing step is performed $\eta_1$ times before restriction and $\eta_2$ times after interpolation. Thus, the computation cost on a seven-point stencil can be approximated by
 \begin{equation}
 T_{\textnormal{comp},S} = 7 \eta \cdot \Bigg(  \sum^{\floor{\log_{\gamma^3} \frac{N}{P}}}_{i=0} \frac{N}{\gamma^{3i}P} + \sum^{\floor{\log_{\gamma^3} N}}_{i=\floor{\log_{\gamma^3} \frac{N}{P}}+1} 1 \Bigg) \cdot t_c ,
 \end{equation}
where the number of smoothing phases $\eta = \eta_1+\eta_2$. In our analysis we assume one smoothing step before restricting and one smoothing step after interpolation.

\subsubsection{Memory Access Costs}

A cache oblivious algorithm for 3-D stencil computations incurs at most $\mathcal{O}(N/Z^{1/3})$ cache misses for each transferring line of size $L$~\cite{Frigo2005}. This number of cache misses matches the lower bound of Hong and Kung~\cite{Wei1981} within a constant factor. We apply this bound to the stencil computations within the multigrid method. Therefore, the memory access cost of smoothing can be represented by
 \begin{equation}
 T_{\textnormal{mem},S} = 7 \eta \cdot \Bigg(  \sum^{\floor{\log_{\gamma^3} \frac{N}{P}}}_{i=0} \frac{ N}{\gamma^{3i}P} + \sum^{\floor{\log_{\gamma^3} N}}_{i=\floor{\log_{\gamma^3} \frac{N}{P}}+1} 1 \Bigg) \frac{L}{ Z^{1/3}} \cdot \beta_{\textnormal{mem}}.
 \end{equation}

\subsubsection{Network Communication Costs}

Communication within the V-cycle takes the form of nearest-neighbor halo exchanges.
In the 3-D multigrid, each processor communicates with its six neighbors where the amount of data exchanged decreases by a factor of $c^{2}$ on each subsequent level. Thus, the communication time at level $l$ is given by
\begin{equation}
 T^l_{\textnormal{net},S} =  T^l_{\textnormal{net},R} =  T^l_{\textnormal{net},I} = 6 \alpha +  \frac{6(N/P)^{2/3}}{\gamma^{2l}} \cdot  \beta_{\textnormal{link}}.
 \end{equation}

\subsection{Model Interpretation}

\subsubsection{Roofline Model}

Arithmetic intensity (AI) is the ratio of total floating-point operations (FLOPs) to total data movement (Bytes). Applications with low arithmetic intensity are typically memory-bound. This means their execution time is limited by the speed at which data can be moved rather than the speed at which computations can be performed, as in compute-bound applications. Hence, memory-bound applications achieve only a small percentage of the theoretical peak performance of the underlying hardware.

The roofline model can be used to assess the quality of attained floating-point performance (GFLOP/s) by combining machine peak performance, machine sustained bandwidth, and arithmetic intensity as follows
\begin{equation}
\textrm{Attainable (GFLOP/s)} = \min (\textrm{Peak (GFLOP/s)}, \textrm{Memory BW} \times \textrm{AI}).
\label{eq:roofline}
\end{equation}
\begin{figure}
\centering
	\includegraphics[width=0.7\textwidth]{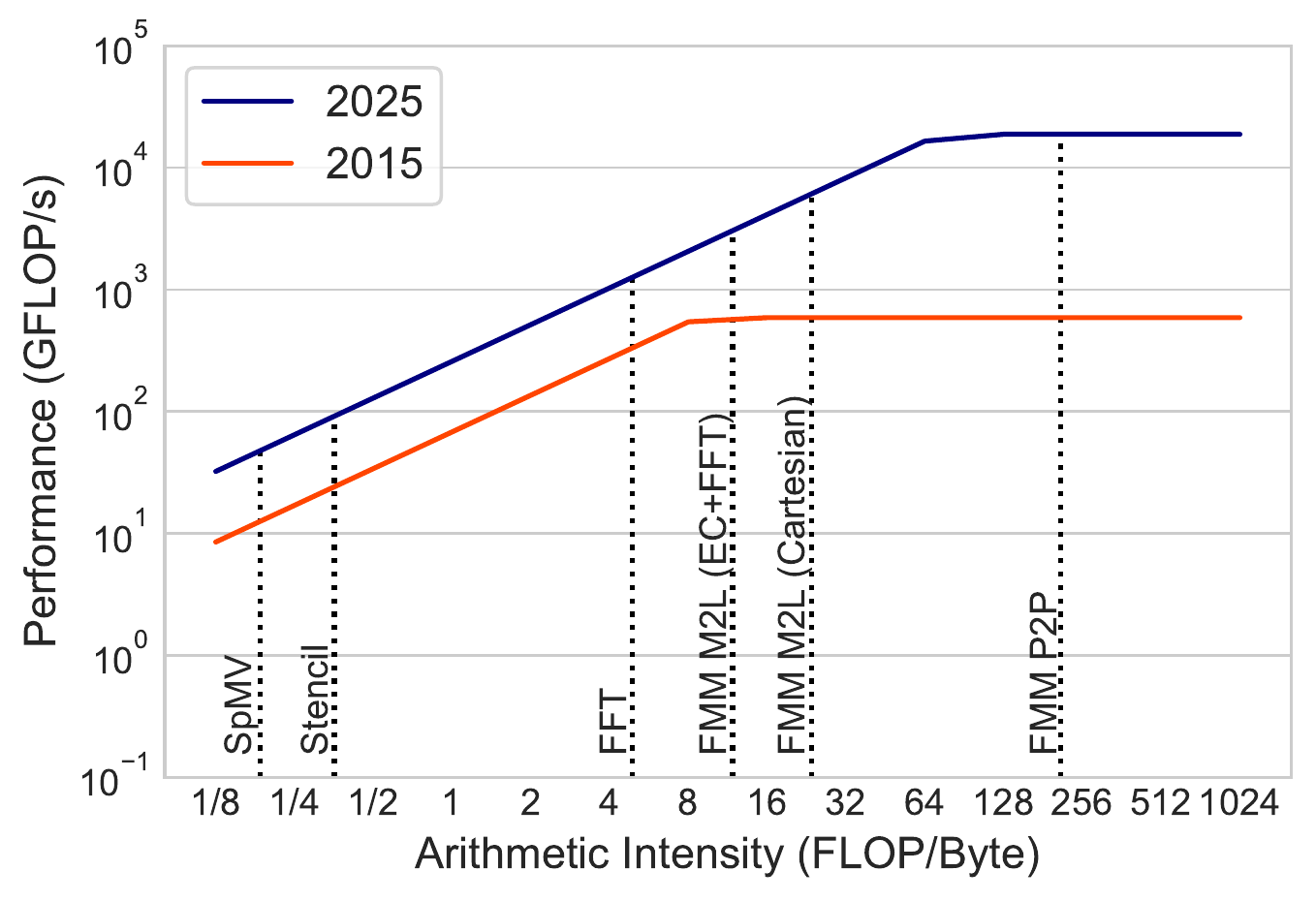}
    \caption{Roofline model and computation intensity of various phases of the FFT, FMM, and MG methods with $N=(32$K$)^3$.}
    \label{fig:roofline}
\end{figure}
Figure~\ref{fig:roofline} shows a roofline model along with the arithmetic intensity of various phases of the FFT, FMM, and MG methods. The ridge point on the roofline model is the processor balance point. All intensities to the left of the balance point are memory bound, whereas all to the right are compute bound. Comparing the three methods shows that the FMM computations have higher arithmetic intensity due to its matrix-free nature. On the other hand, SpMV and stencil operations, which are the basic building blocks of the classic algebraic and geometric multigrid methods, have low arithmetic intensities. The 3-D FFT has an intermediate arithmetic intensity that grows slowly with the problem size.

\begin{figure}
\centering
	\includegraphics[width=0.7\textwidth]{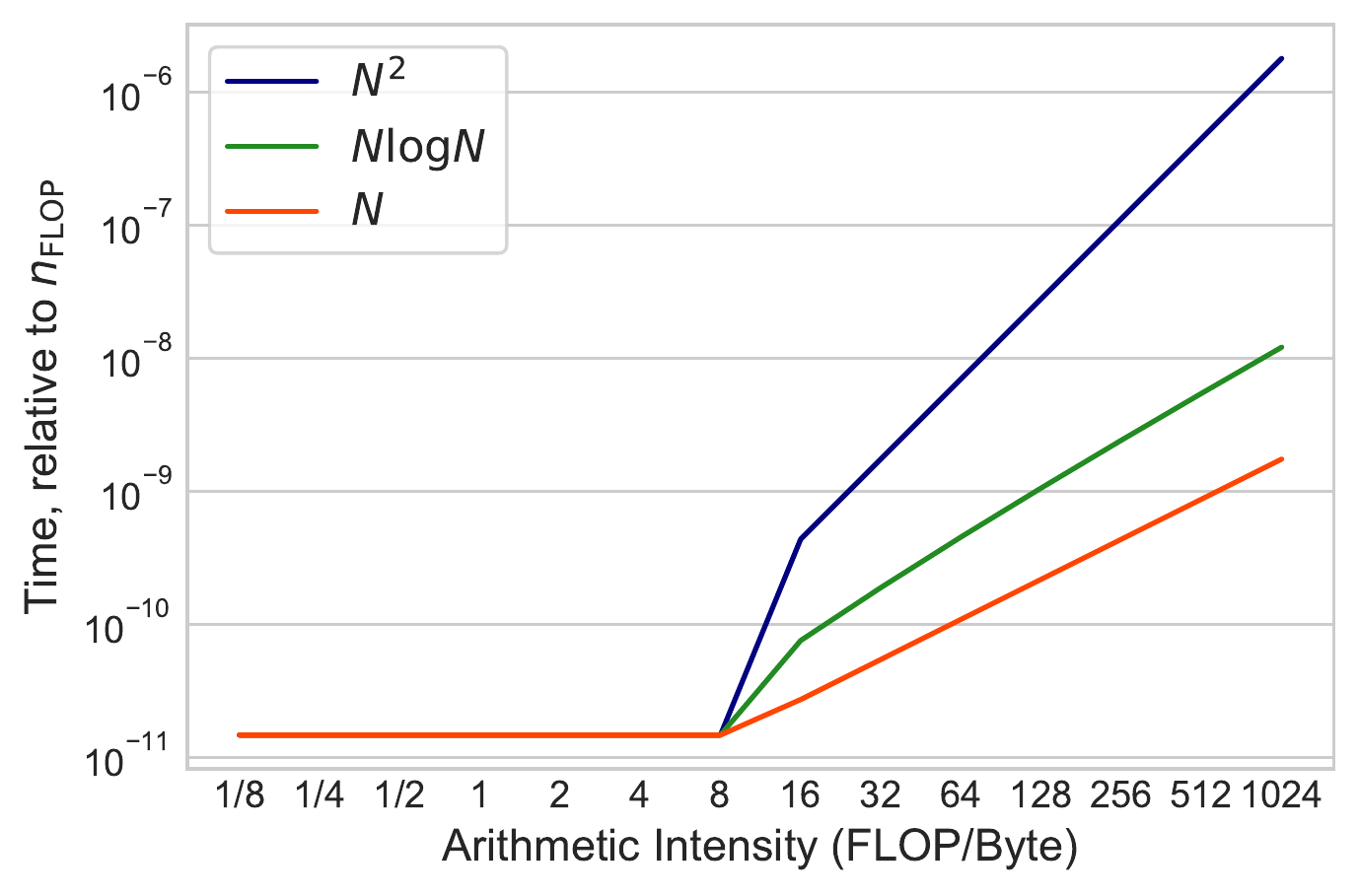}
    \caption{Execution time (normalized to $n_{\textnormal{FLOP}}$) for various computational complexities.}\label{fig:efficiencyT}
\end{figure}
In order to understand the relation between computation time and arithmetic intensity, the algorithmic efficiency must be taken into account. Using~\eqref{eq:T}, Figure~\ref{fig:efficiencyT} shows that computation time is independent of the algorithmic computational complexity up until the processor balance point. Beyond that point, it increases as a function of the algorithmic complexity.

\subsubsection{Projecting Forward}

Figure~\ref{fig:roofline} also shows the roofline model of a possible future CPU processor. The characteristics of the processor are based on extrapolating historical technology trends. These trends are summarized in Table~\ref{t:exa}. From Figure~\ref{fig:roofline} we observe that although FMM is compute-bound on contemporary systems, it could become memory-bound at exascale.

\section{Resiliency}\label{sec:resiliency}

The focus of this section is on the main memory and network errors. Therefore, to assess the resilience of FFT, FMM, and MG, we quantify the vulnerability of the data structures and communication patterns used within these methods.

For data structures, we use the data vulnerability factor ($DVF$) introduced in~\cite{Yu2014}. The $DVF$ for a specific data structure ($DVF_d$) is defined as
\begin{equation}
DVF_d = FIT \times T_{\textnormal{exe}} \times S_d \times N_{ha},
\end{equation}
where $FIT$ is the failure in time (failures per billion hours per Mbit), $T_{\textnormal{exe}}$ is the application execution time, $S_d$ is the size of the data structure, and $N_{ha}$ is the number of accesses to the hardware (main memory in this study). To estimate $N_{ha}$, we use the number of cache misses approximated for FFT, FMM, and MG in Section~\ref{sec:concurrency}.

The DVF of an application ($DVF_a$) can be calculated by
\begin{equation}
DVF_a = \sum^D_{i=1} DVF_{d_i},
\end{equation}
where $D$ is the number of major data structures in the application.

For communication, we introduce the communication vulnerability factor ($CVF$) which reflects the communication pattern and network characteristics. The $CVF$ for a specific kernel ($CVF_k$) is defined as
\begin{equation}
CVF_k = m \times T_{\textnormal{net}} \times RF_n,
\end{equation}
where $m$ is the maximum number of messages sent, $T_{\textnormal{net}}$ is the application communication time, and $RF_n$ is the network resilience factor defined for uniform deterministic traffic~\cite{Liu2009} as follows
\begin{equation}
RF_n = \bar{h} p_e + b p_b^2,
\end{equation}
where $\bar{h}$ is the average route length and $p_e$ is the effective link failure probability given by
\begin{equation}
p_e = a p_a + 2 b p_b - b p_b^2,
\end{equation}
where $a$ is the probability of the occurrence of an event $A$ that can only affect the status of a link, $b$ is the probability of the occurrence of an event $B$ that can affect the status of all the links incident at a node, and $p_a$ and $p_b$ are the probability that events $A$ and $B$, respectively, can lead to link failure. Here, we define $\bar{h}$ as the diameter of the network formed by $P$ nodes.

For multilevel methods, the $CVF$ is calculated at each level individually. Thus, the application $CVF_a$ is given by
\begin{equation}
CVF_a = \sum^K_{k=1} CVF_{k},
\end{equation}
where $K$ is the number of key kernels in the application and $CVF_{k}$ is given by
\begin{equation}
CVF_k = \sum^L_{l=1} CVF^l_{k},
\end{equation}
where $L$ is the number of levels.

\subsection{Model Interpretation}

\subsubsection{Data Structures Vulnerability}

\begin{table}[t]
\centering
\caption{Data vulnerability factors of FFT, FMM, and MG.} \label{t:dvf}
\small
\begin{tabular}{ccc}
 \toprule
  & 2015 & 2025 \\ [0.5ex]
 \midrule
FFT & 0.003 & 0.431 \\
FMM & 0.376 & 41.20 \\
MG & 0.037 & 2.101 \\
\bottomrule
\end{tabular}
\end{table}
%
Table~\ref{t:dvf} shows the DVF of the FFT, FMM, and MG methods. FMM has random memory access pattern as memory accesses to the tree are random. FFT and MG, on the other had, have template-based memory access pattern where accesses to elements of the data structure mesh follow specific topology or stencil information instead of arbitrarily constructed. Table~\ref{t:dvf} shows that algorithms with random memory access pattern, such as FMM, have higher DVF than algorithms with template-based memory access pattern, such as FFT and MG. Therefore, the FMM data structures are more sensitive to memory errors compared to FFT and MG. These observations are consistent with the ones in~\cite{Yu2014}.

\subsubsection{Communication Vulnerability}

\begin{table}[t]
\centering
\caption{Communication vulnerability factors of FFT, FMM, and MG.} \label{t:cvf}
\small
\begin{tabular}{ccc}
 \toprule
  & 2015 & 2025 \\ [0.5ex]
 \midrule
FFT & 0.020 & 2.934 \\
FMM & 3.4e-4 & 0.004 \\
MG & 1.2e-5 & 4.9e-5 \\
\bottomrule
\end{tabular}
\end{table}
The communication vulnerability factors of FFT, FMM, and MG are shown in Table~\ref{t:cvf}. As expected, the all-to-all communication pattern of FFT makes it more sensitive to network failures compared to the hierarchical methods, FMM and MG. The hierarchical nature of FMM and MG reduces the $\mathcal{O}(\sqrt{P})$ communication complexity of FFT to $\mathcal{O}(\log{P})$. This communication complexity is likely to be optimal for elliptic problems, since an appropriately coarsened representation of a local forcing must somehow arrive at all other parts of the domain for the elliptic equation to converge.

\subsubsection{Projecting Forward}

For exascale projections, we scale the problem size by the ``Cores" 10-year increase factor from Table~\ref{t:exa}. In Tables~\ref{t:dvf} and~\ref{t:cvf}, the problem size per processor is $N/P=32^3$ for 2015 and $N/P={65}^3$ for 2025. We also scale the number of processes by the ``Processors" increase factor from $P=11,889$ to $P=372k$ and assume that the effective link failure probability $p_e$ remains constant over time. The results show that both the DVF and CVF are expected to increase on future exascale systems with more drastic increase in the DVF.

\section{Heterogeneity}
\label{sec:heterogeneity}
In this section we adapt the FFT, FMM, and MG execution models introduced in Section~\ref{sec:concurrency} to accelerators. In particular, we consider NVIDIA GPUs. One of the main architectural differences between GPUs and CPUs is the relatively small caches on GPUs which makes reusing data in the fast memory more difficult. Another bottleneck to consider on heterogeneous systems is the the PCIe bus. Due to the high compute capability of the GPU, the PCIe bus can have a significant impact on performance. The PCle transfer time for $n$ elements is given by
\begin{equation}
T_{\textnormal{PCIe}}(n) = n\beta_{\textnormal{PCIe}},
\end{equation}
where $\beta_{\textnormal{PCIe}}$ is the I/O bus bandwidth inverse in seconds per element. For FFT, FMM, and MG, the PCle transfer time is given by
\begin{equation}
T_{\textnormal{PCIe}} = 2 \times \frac{N}{P} \beta_{\textnormal{PCIe}},
\end{equation}
where the factor of two accounts for the two ways transfer. Here, we assume that each processor has a direct network connection, optimistically avoiding PCIe channels.

\subsection{Model Interpretation}

\subsubsection{Roofline Model}

\begin{figure}
\centering
	\includegraphics[width=0.7\textwidth]{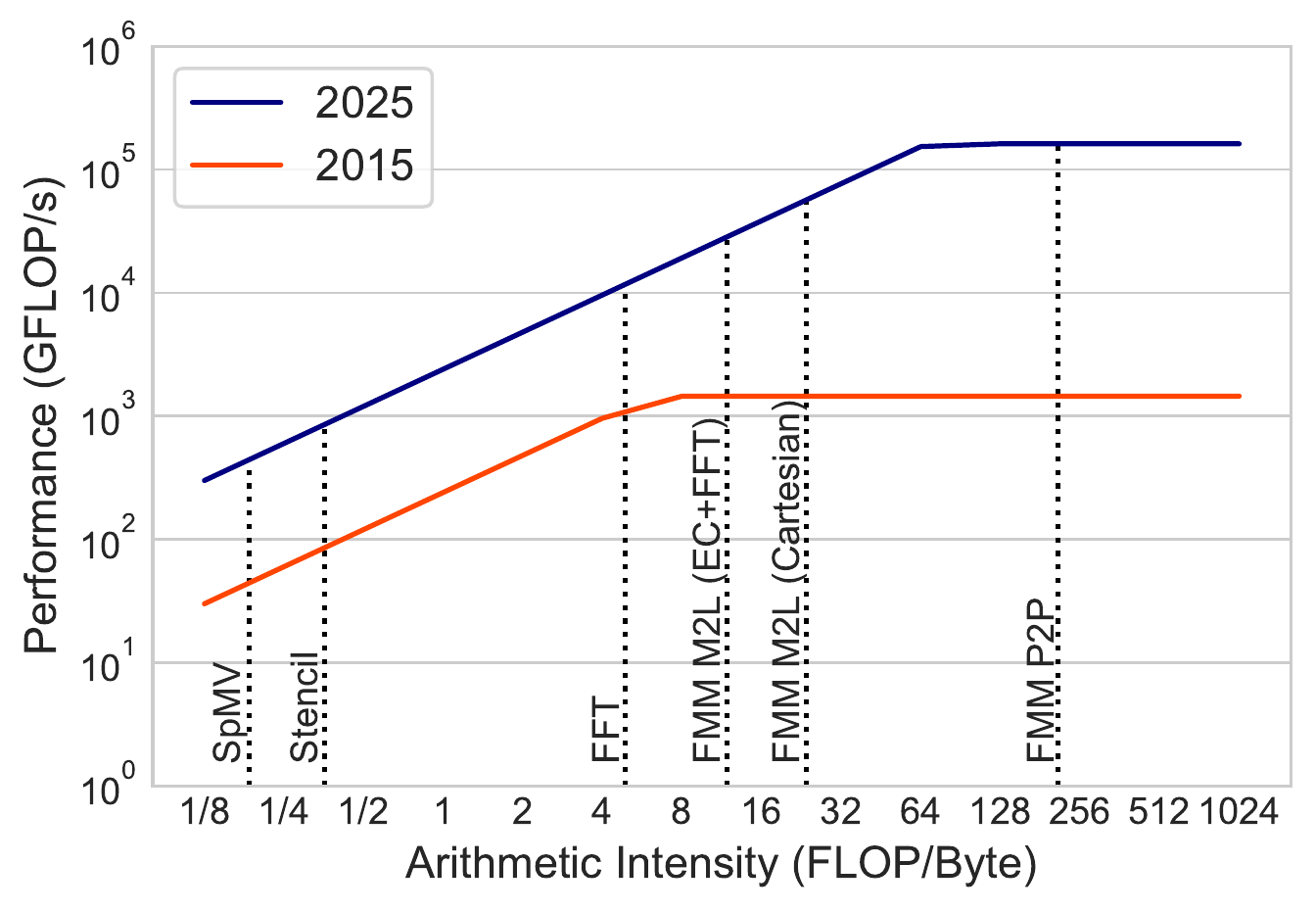}
    \caption{Roofline model of NVIDIA Tesla GPU and computation intensity of various phases of the FFT, FMM, and MG methods with $N=(32$K$)^3$.}
    \label{fig:rooflineGPU}
\end{figure}
Figure~\ref{fig:rooflineGPU} shows roofline models of NVIDIA Tesla GPU and of a possible future GPU processor that is based on extrapolating historical technology trends. Similar to the CPU exascale projection results, Figure~\ref{fig:rooflineGPU} shows that kernels that are compute-bound on contemporary systems could become memory-bound at exascale.

\subsubsection{Projecting Forward}

Using the analytical models, we predict the communication time of FFT, FFM, and MG for large-scale problems on possible future GPU-only and CPU-only exascale systems. The machine characteristics of the exascale systems are based on the trends summarized in Table~\ref{t:exa}.

\begin{figure}
\centering
	\begin{subfigure}[t]{0.311\textwidth}
        \centering
        \includegraphics[width=\textwidth]{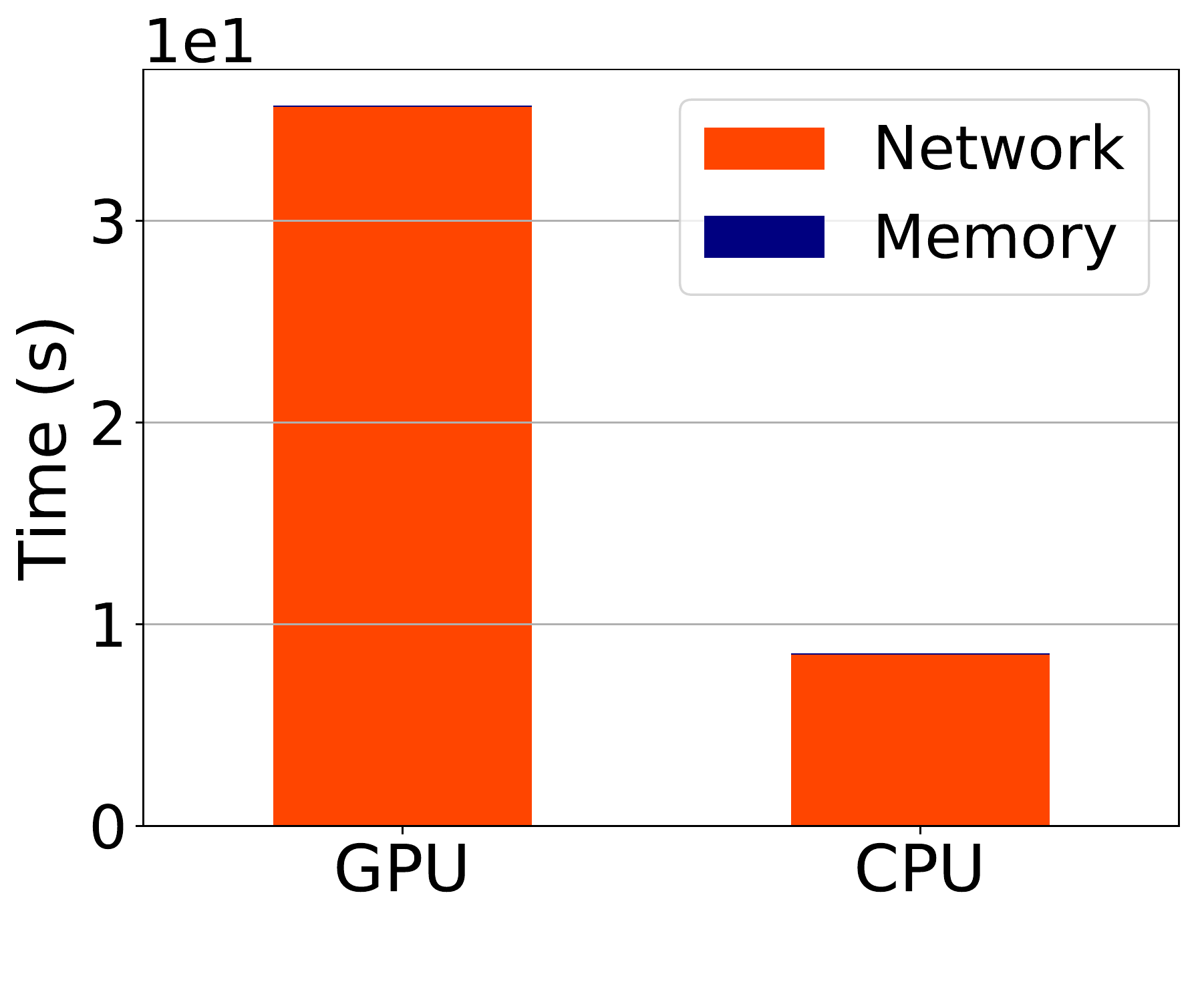}
        \caption{FFT}
    \end{subfigure}
	\begin{subfigure}[t]{0.325\textwidth}
        \centering
        \includegraphics[width=\textwidth]{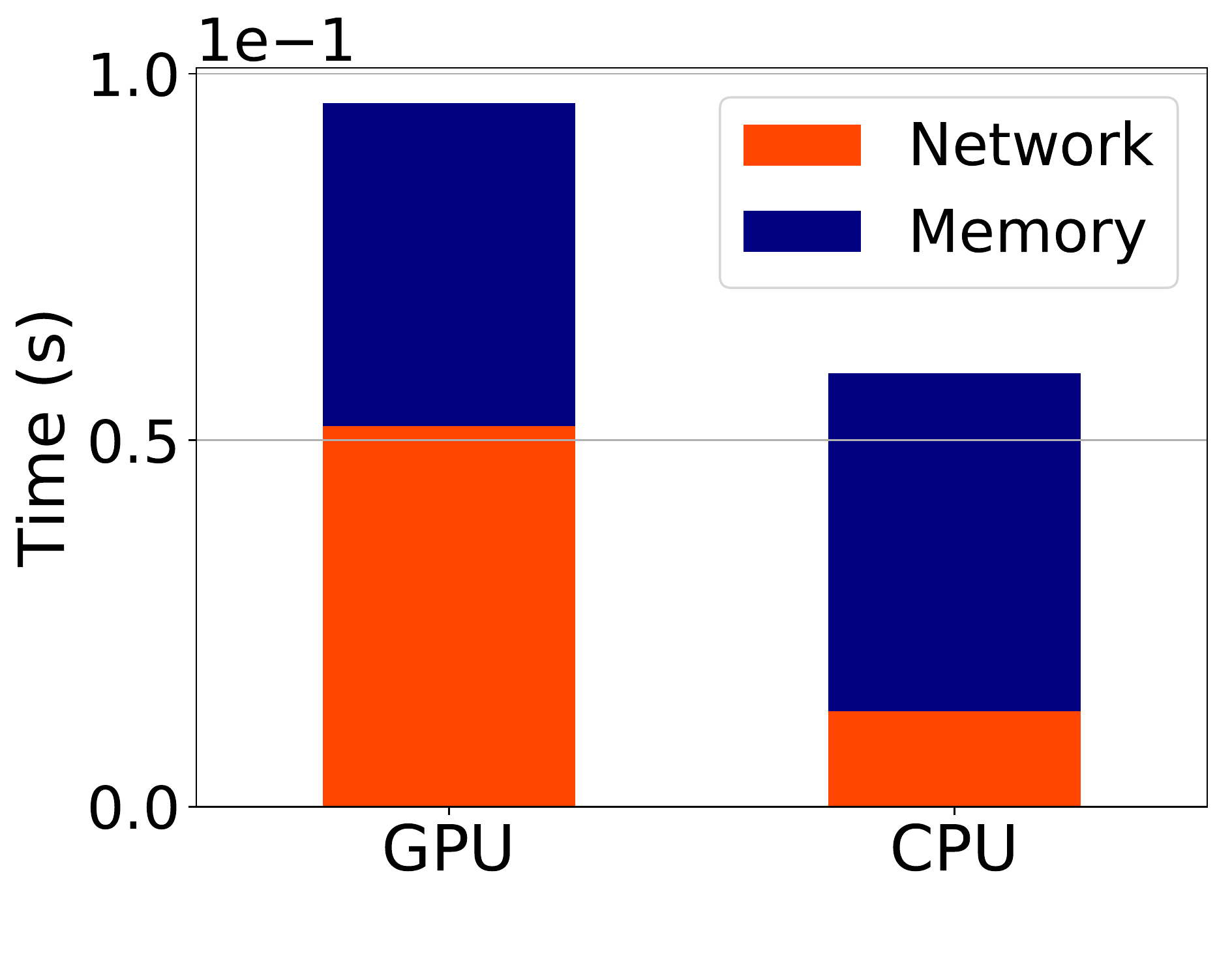}
        \caption{FMM}
    \end{subfigure}
	\begin{subfigure}[t]{0.31\textwidth}
        \centering
        \includegraphics[width=\textwidth]{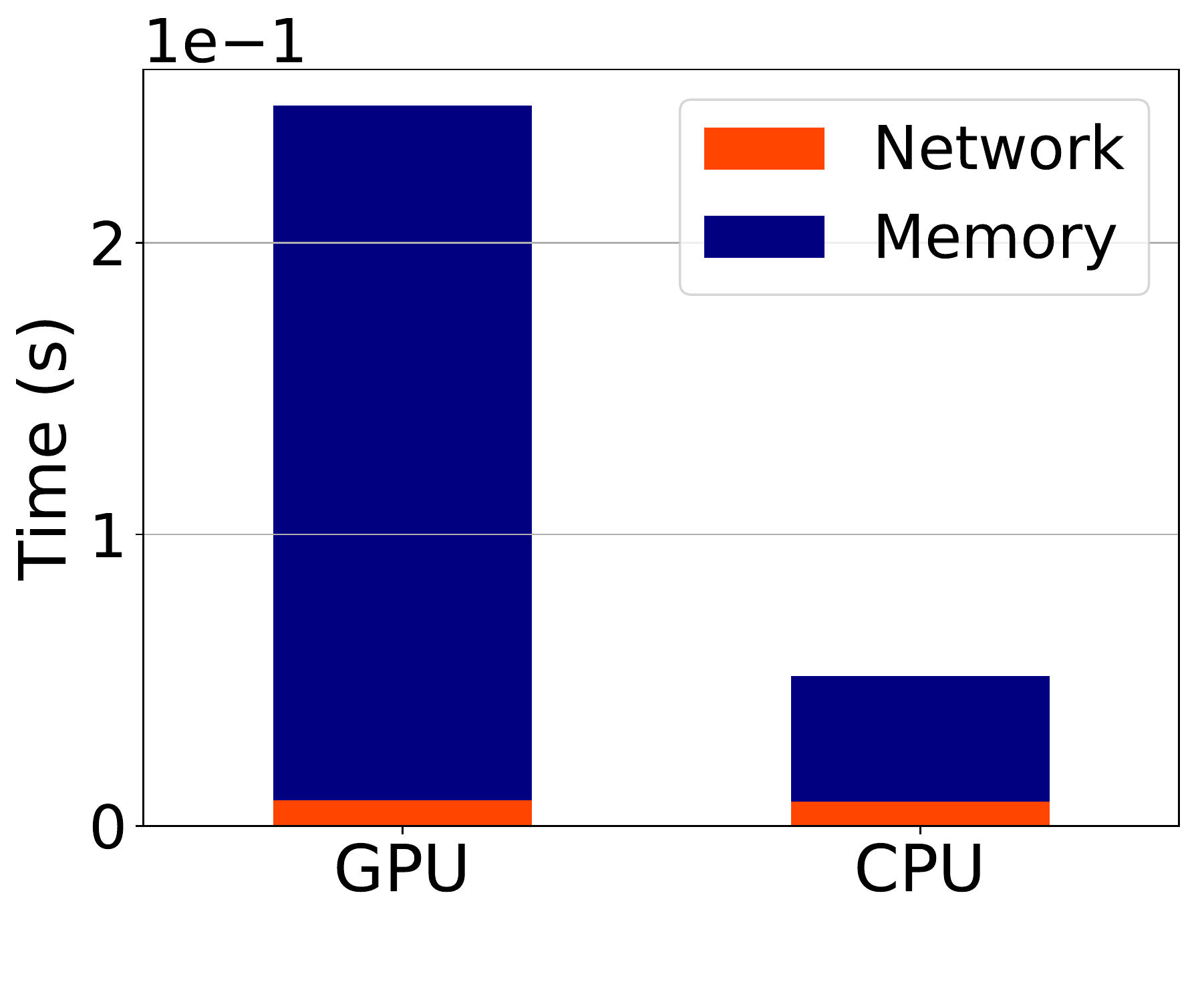}
        \caption{Multigrid}
    \end{subfigure}
    \caption{Exascale projections of the FFT, FMM, and MG methods on GPU- and CPU-only systems with $N=(65$K$)^3$.}
    \label{fig:gpu}
\end{figure}
Figure~\ref{fig:gpu} shows the communication time of FFT, FMM, and MG split into memory and network access costs. We consider extrapolated GPU-only and CPU-only systems that have the same peak performance of $7$ exaFLOPS. The GPU-only system has $43.3$K processors while the CPU-only system has $372$K processors. The results show that all methods spend less time in both forms of communication on the CPU-only system. However, this system requires almost $8.7 \times$  as many processors as the GPU-only system. This cost could be prohibitive. 

Figure~\ref{fig:gpu} shows that the FFT communication time is dominated by the network access cost which is expected given the all-to-all communication pattern of FFT. On the other hand, memory access cost dominates MG communication time. This implies that intra-node communication could become the bottleneck that limits the scalability of MG on exascale systems.

\section{Energy}\label{sec:energy}

To characterize power and energy efficiency of the FFT, FMM, and MG methods, we use the energy roofline model introduced in~\cite{Choi2013}. This model bounds power consumption as a function of the total floating-point operations and total amount of data moved. The energy cost (Joules) is defined by
\begin{equation}
  E \equiv E_{\!_{\textnormal{FLOP}}} + E_{\!_{\textnormal{mem}}} + E_0(T_{\textnormal{exe}}),
\label{eq:E1}
\end{equation}
where $E_{\!_{\textnormal{FLOP}}}$ is the total energy consumption of the computation, $E_{\!_{\textnormal{mem}}}$ is the total energy consumption of memory traffic, and $E_0$ is a measure of energy leakage as a function of execution time, $T_{\textnormal{exe}}$.

Suppose the energy cost is linear in $T_{\textnormal{exe}}$ with a fixed constant power $\pi_0$,~\eqref{eq:E1} can be written as
\begin{equation}
E = n_{\textnormal{FLOP}} \cdot \epsilon_{\!_{\textnormal{FLOP}}} + n_{\textnormal{mem}} \cdot \epsilon_{\!_{\textnormal{mem}}}  + \pi_0 \cdot T_{\textnormal{exe}},
\end{equation}
where $n_{\textnormal{FLOP}}$ is the number of FLOPs, $n_{\textnormal{mem}}$ is the number of main memory operations, and $\epsilon_{\!_{\textnormal{FLOP}}}$ and $\epsilon_{\!_{\textnormal{mem}}}$ are fixed energy per computation and per memory operation, respectively. Defining the energy balance $B_{\epsilon} \equiv \epsilon_{\!_{\textnormal{mem}}} /\epsilon_{\!_{\textnormal{FLOP}}}$, the above equation becomes
\begin{equation}
E = n_{\textnormal{FLOP}} \cdot \epsilon_{\!_{\textnormal{FLOP}}} \cdot ( 1 + \frac{B_{\epsilon}}{AI} + \frac{\pi_0 \cdot T_{\textnormal{exe}}}{\epsilon_{\!_{\textnormal{FLOP}}} \cdot n_{\textnormal{FLOP}}}),
\label{eq:E}
\end{equation}

\begin{figure}
\centering
	\includegraphics[width=0.7\textwidth]{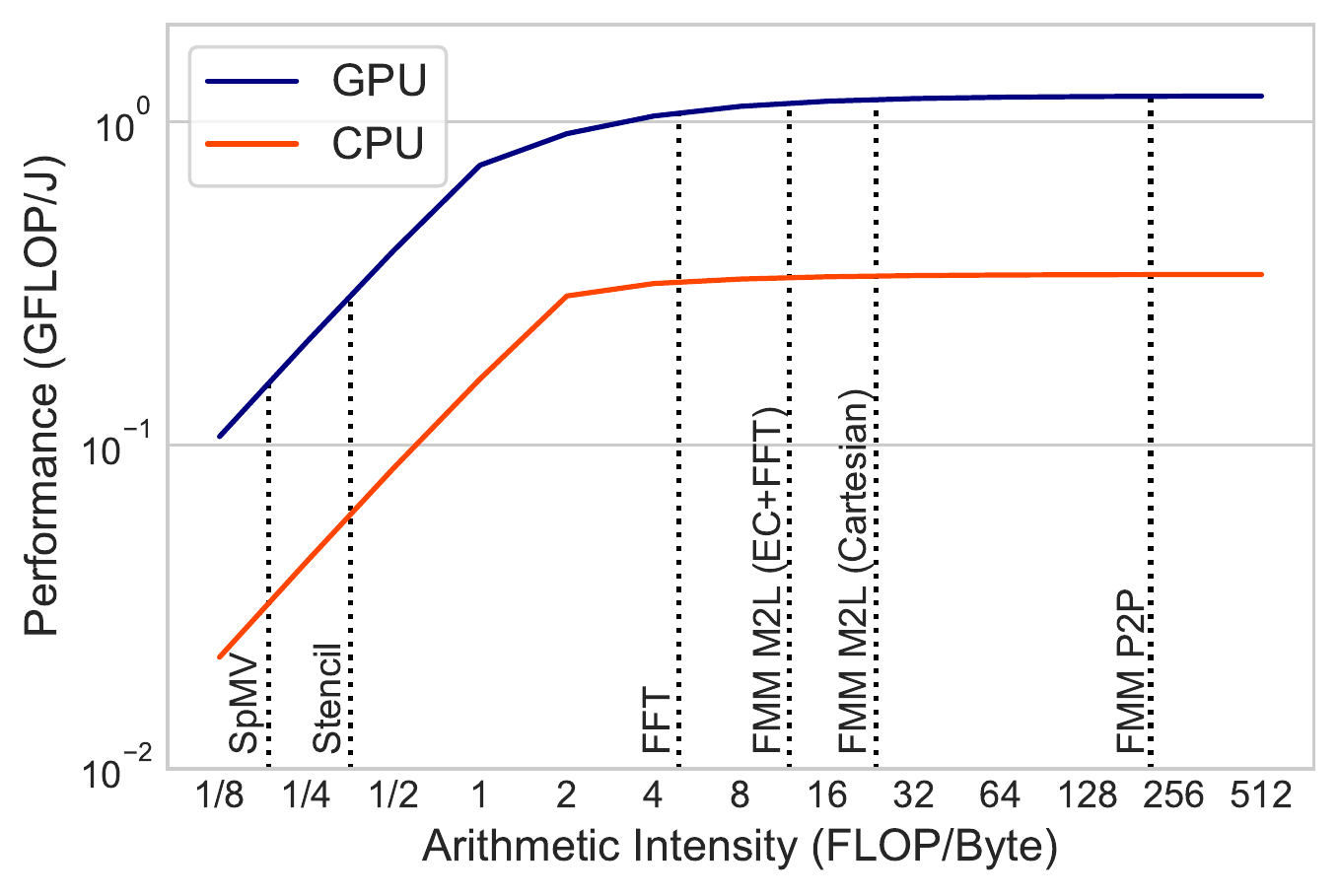}
    \caption{Energy roofline model of NVIDIA GTX 580 GPU and Intel i7-950 (double-precision) along with the computation intensity of various phases of FFT, FMM, and MG (GPU: $\epsilon_{\!_{\textnormal{FLOP}}} \approx 212$ pJ per FLOP, $\epsilon_{\!_{\textnormal{mem}}}
\approx 513$ pJ per Byte, $\pi_0 \approx 122$ Watts, $t_c \approx 5.1$ ps per FLOP, and $\beta_{\textnormal{mem}} \approx 5.2$ ps per Byte; CPU: $\epsilon_{\!_{\textnormal{FLOP}}} \approx 670$ pJ per FLOP, $\epsilon_{\!_{\textnormal{mem}}}
\approx 795$ pJ per Byte, $\pi_0 \approx 122$ Watts, $t_c \approx 18$ ps per FLOP, and $\beta_{\textnormal{mem}} \approx 39$ ps per Byte~\cite{Choi2014power}).}\label{fig:rooflineE}
\end{figure}
\begin{figure}
\centering
	\includegraphics[width=0.7\textwidth]{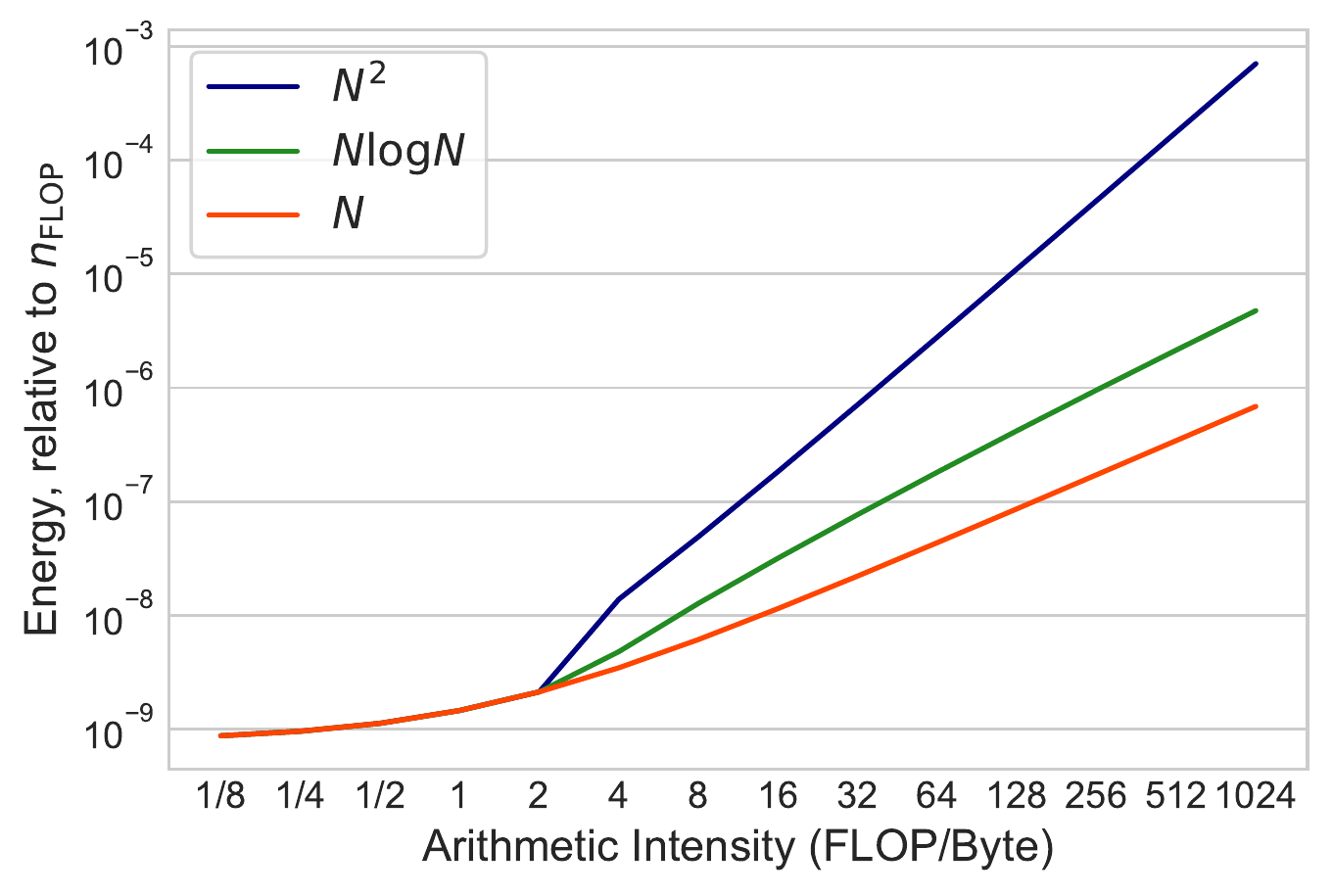}
    \caption{Energy consumption (normalized to $n_{\textnormal{FLOP}}$) for various computational complexities.}\label{fig:efficiencyE}
\end{figure}
Using~\eqref{eq:E}, Figure~\ref{fig:rooflineE} shows the energy roofline model along with the arithmetic intensity of various phases of FFT, FMM, and MG\@. The figure shows that algorithms with higher arithmetic intensity have better energy efficiency. However, total energy consumption depends heavily on the algorithmic efficiency. Similar to Figure~\ref{fig:efficiencyT}, Figure~\ref{fig:efficiencyE} shows that total energy consumption is independent of the algorithmic computational complexity for all intensities to the left of the processor balance point whereas energy consumption increases as a function of the algorithmic complexity beyond that point.


\subsection{Projecting Forward}
%

The classic equation for dynamic power is given by
\begin{equation}\label{eq:PDyn}
  \textnormal{Power}_{\textnormal{dyn}} = \alpha C V^2 f,
\end{equation}
where $C$ is the load capacitance, a physical property of the material, $V$ is the supply voltage, and $f$ is the clock frequency. Hence, $\epsilon_{\!_{\textnormal{FLOP}}}$ and $\epsilon_{\!_{\textnormal{mem}}}$ become
\begin{equation}\label{eq:EfDyn}
  \epsilon_{\!_{\textnormal{FLOP}}} = t_c \cdot \textnormal{Power}_{\textnormal{dyn}},
\end{equation}
and
\begin{equation}\label{eq:EmDyn}
  \epsilon_{\!_{\textnormal{mem}}} = \beta_{\textnormal{mem}} \cdot \textnormal{Power}_{\textnormal{dyn}}.
\end{equation}

\begin{table}
\centering
\caption{Component scaling with node size.  Ratios are given in reference to 45 nm~\cite{Esmaeilzadeh2013}.}\label{t:power}
\small
\begin{tabular}{ccccc}
 \toprule
 Tech node (nm) & Frequency & Voltage & Capacitance & Power\\
 \midrule
45 & 1.00 & 1.00 & 1.00 & 1.00 \\
32 & 1.10 & 0.93 & 0.75 & 0.71 \\
22 & 1.19 & 0.88 & 0.56 & 0.52 \\
16 & 1.25 & 0.86 & 0.42 & 0.39 \\
11 & 1.30 & 0.84 & 0.32 & 0.29 \\
8 & 1.34 & 0.84 & 0.24 & 0.22 \\
\bottomrule
\end{tabular}
\end{table}
\begin{figure}
\centering
	\includegraphics[width=0.7\textwidth]{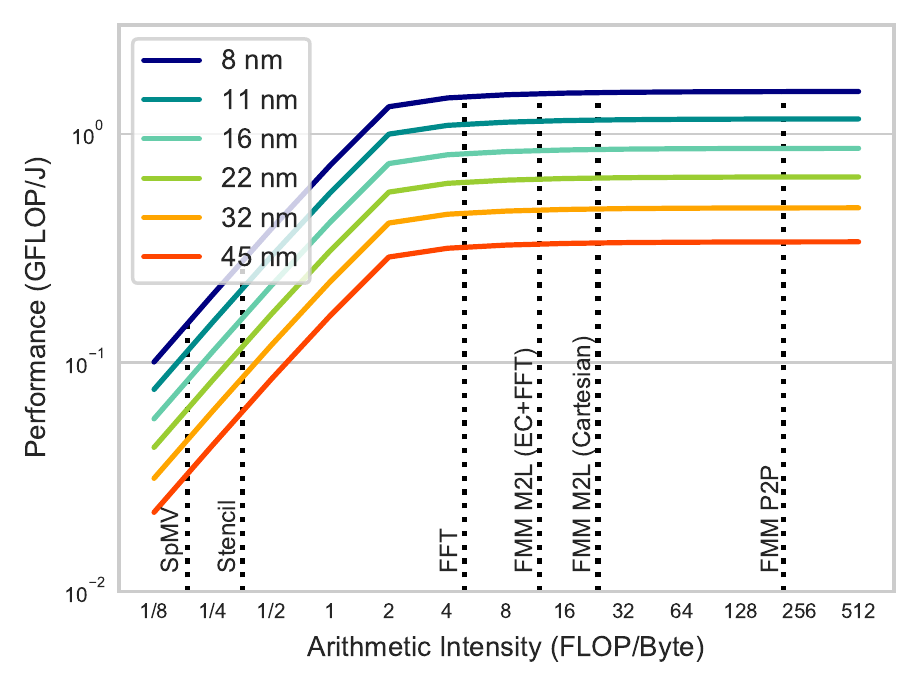}
    \caption{Energy roofline model of various technology nodes along with the computation intensity of FFT, FMM, and MG.}\label{fig:projectE}
\end{figure}
To estimate the energy efficiency of future multicore chips, we use the transistor scaling projection model presented in~\cite{Esmaeilzadeh2013}. This model provides the area, voltage, and frequency scaling factors for technology nodes from 45nm through 8nm. These factors are summarized in Table~\ref{t:power}. Figure~\ref{fig:projectE} shows how the energy efficiency is predicted to improve as the node size decreases. Nevertheless, the per-transistor power efficiency improvements have slowed in comparison to the historic rates. Microarchitecture innovations are needed to improve the energy efficiency.

\section{Memory}
\label{sec:memory}

Memory hierarchy is expected to change at exascale based on both new packaging capabilities and new technologies to provide the required bandwidth and capacity. Local RAM and non-volatile-memory (NVRAM) will be available either on or very close to the nodes to reduce wire delay and power consumption. One of the leading proposed mechanism to emerge in the memory hierarchy is the 3-D stacked memory which enables DRAM devices with much higher bandwidths than traditional DIMMs (dual in-line memory module).

\begin{table}
\caption{Approximate bandwidths and capacities of memory subsystem.}
\centering
\small
\begin{tabular}{cccc}
 \toprule
& Configuration & Bandwidth & Capacity\\ [0.5ex]
 \midrule
Single-Level & HMC & 240 GB/s per stack & 16 GB per stack\\
 \midrule
\multirow{2}{*}{Multi-Level DRAM} & HBM & 200 GB/s per stack & 16 GB per stack \\
& DDR & 20 GB/s per channel & 64 GB per DIMM \\
 \midrule
High Capacity Memory & NVRAM & 10 GB/s & $4 - 8$ $\times$ DRAM\\
 \bottomrule
\end{tabular}
\label{t:memory}
\end{table}

Deeper memory hierarchy is expected at exascale with each level composed of a different memory technology. One proposed memory hierarchy for exascale systems consists of~\cite{Ang2014}: a high-bandwidth 3-D stacked memory, such as high bandwidth memory (HBM) standard or hybrid memory cube (HMC) technology, a standard DRAM, and NVRAM memory. Approximate bandwidths and capacities of the proposed memory subsystem are shown in Table~\ref{t:memory}.

\begin{figure}
\centering
	\includegraphics[width=0.7\textwidth]{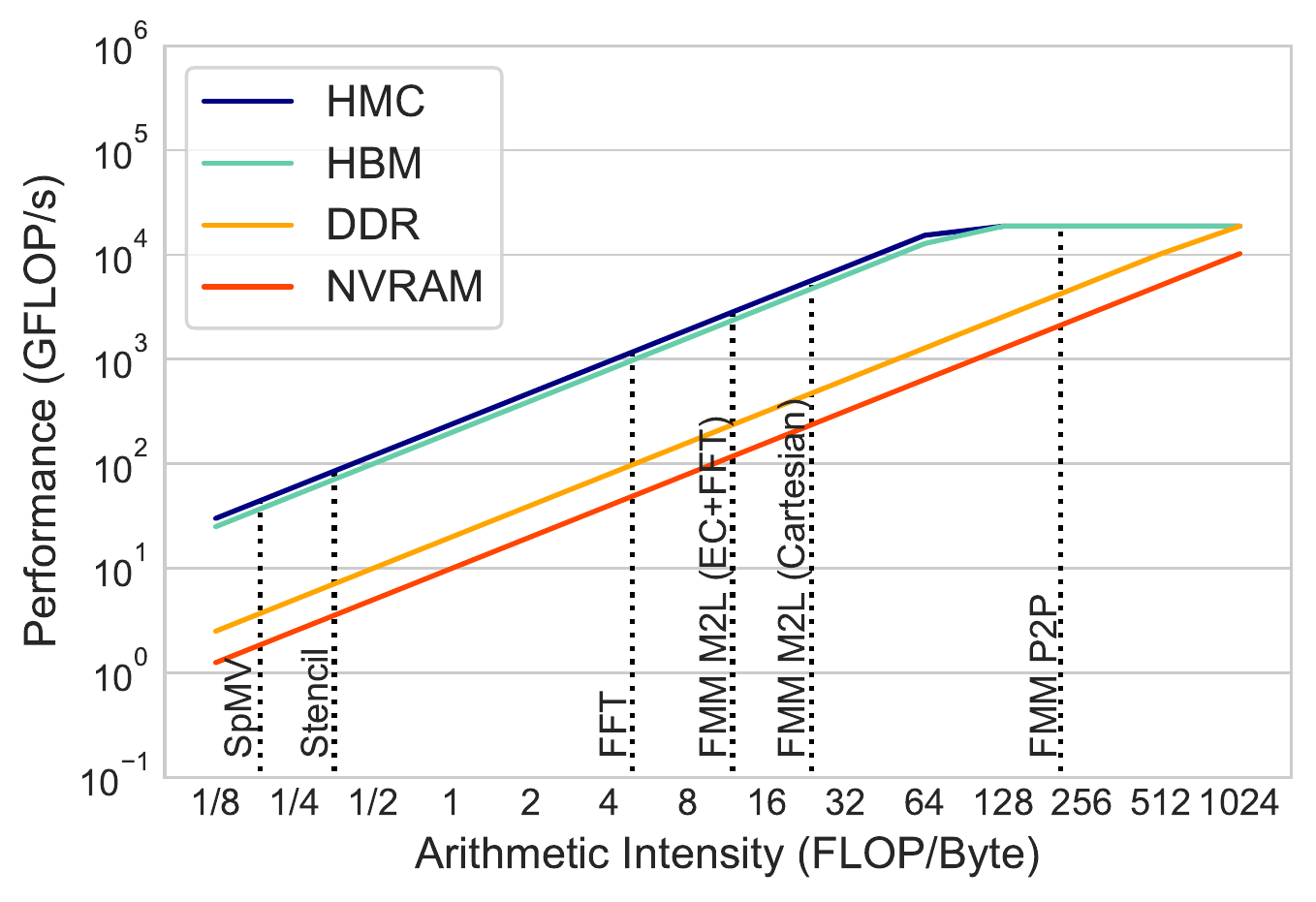}
    \caption{Memory-aware roofline model of a possible exascale machine along with the computation intensity of various phases of FFT, FMM, and multigrid methods with $N=(32$K$)^3$.}
    \label{fig:rooflineMemory}
\end{figure}
Figure~\ref{fig:rooflineMemory} shows memory-aware roofline models of the different memory technologies. The roofline models can be derived by substituting the bandwidth values from Table~\ref{t:memory} into~\eqref{eq:roofline}. Emerging 3-D stacked DRAM devices, such as HBM and HMC, will significantly increase available memory bandwidth. However, with the exponential increase in core counts, stacked DRAM will only move the memory wall and is unlikely to break through it~\cite{Radulovic2015}. Figure~\ref{fig:rooflineMemory} also shows the large difference in attainable performance between different levels of the memory hierarchy. Exascale applications need to exploit data locality and explicitly manage data movement to minimize the cost of memory accesses and to make the most effective use of available bandwidth.

\section{Observations}
\label{sec:observations}

\begin{itemize}
\item Algorithms that are known to be compute-bound on current architectures, such as the FMM, could become memory-bound on future CPU- and GPU-based exascale systems. 

\item Execution time and energy consumption are independent of the algorithmic computational complexity up until the processor balance point. They increase as a function of the algorithmic complexity beyond that point.

\item Heterogeneous systems are important for energy efficient scientific computing.

\item It is well known that GPUs deliver more peak performance and bandwidth relative to high-end CPUs. This performance gap is likely to increase towards exascale, as shown in Figure~\ref{fig:gap}. 
\begin{figure}[h]
\centering
	\includegraphics[width=0.7\textwidth]{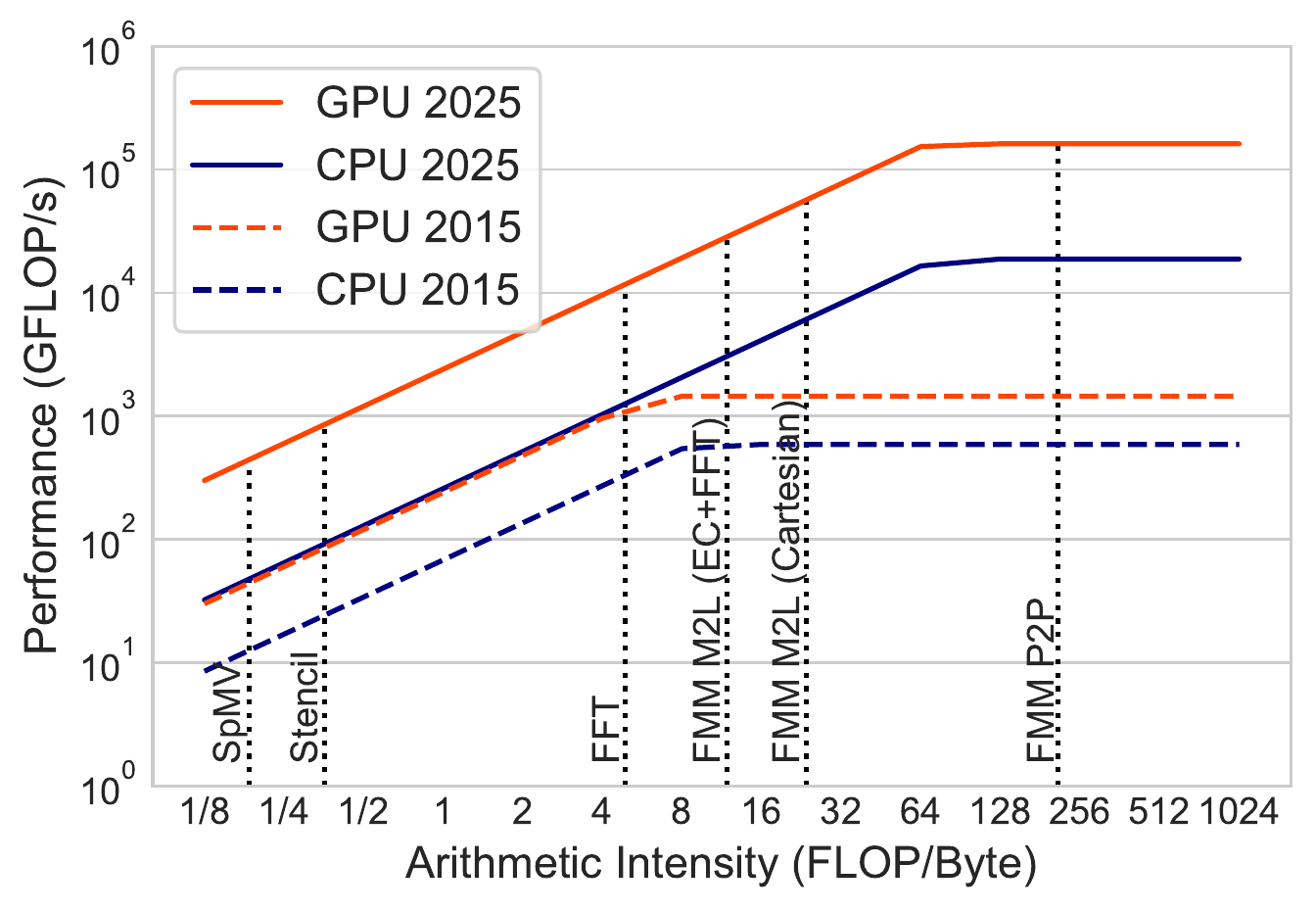}
    \caption{CPU- and GPU-based roofline models.}\label{fig:gap}
\end{figure}


\item Emerging 3-D stacked DRAM devices will significantly increase available memory bandwidth. However, with the exponential increase in core counts, stacked DRAM will only move the memory wall and is unlikely to break through it.

\end{itemize}

\section{Conclusions}
\label{sec:conclusions}

Recent efforts~\cite{Dongarra2011} have identified several constraints in the design of exascale software that include massive concurrency, resilience management, exploiting the high performance of heterogeneous systems, energy efficiency, and utilizing the deeper and more complex memory hierarchy expected at exascale. In this manuscript, we perform model-based comparison of the FFT, FMM, and MG methods  vis-\`a-vis these challenges. We believe that the importance of each of these challenges is application dependent. This paper provides metrics for researchers to quantify these challenges and their importance relative to the applications of interest.

Modeling FFT, FMM, and MG relative to these challenges has contributed to our understanding of the main steps that must be taken on both application and architecture sides to help overcoming these challenges.

On the application side:
\begin{itemize}
\item \textit {Rethink algorithms to reduce memory requirements.} Data movement is the dominant factor that limits performance and efficiency on contemporary architecture. Attainable floating-point performance of memory-bound applications is limited by the memory bandwidth. Furthermore, a significant portion of the energy consumption of modern supercomputers is caused by memory operations. Reducing data movements leads to higher arithmetic intensity, lower memory bandwidth usage, lower energy consumption, and better scalability with the number of cores.

\item \textit {Rethink algorithms to improve arithmetic intensity.} High arithmetic intensity is essential for achieving good performance and efficiency. Possible approaches to increase the arithmetic intensity include improving data locality, combining multiple kernels into a single high arithmetic intensity kernel, and reducing the memory footprint by, for example, using matrix-free approaches as in the FMM.

\item \textit{Design for sustainability.} Resilience is a major obstacle on the road to exascale. Our projections show that resilience is expected to be a much larger issue on exascale systems than it is on current petascale computers. New resilience paradigms are required.

\item \textit {Enable energy-efficient software.} In addition to reducing memory operations and improving arithmetic intensity, power consumption can be reduced at the software side by efficiently exploiting thread level parallelism to ensure balance between performance gains and increases in energy consumption.
\end{itemize}

On the architectural side:
\begin{itemize}
\item \textit {Increasing memory bandwidth.} Radical technology advances are needed to improve local memory bandwidth.

\item \textit {Enable energy-efficient computers.} Ongoing research efforts to improve energy efficiency include~\cite{Getov2016}: dynamic frequency scaling, power-aware applications, energy management throughout the hardware/software stack, and optimization techniques for balancing performance and power. Nevertheless, disruptive technology breakthroughs are still needed to enable energy efficient computers.

\end{itemize}

\section*{Acknowledgments}
The authors would like to thank Prof.~David Keyes (KAUST) for many insightful discussions. This material is based in part upon work supported by the Department of Energy, National Nuclear Security Administration, under Award Number DE-NA0002374.


\bibliography{ref}

\end{document}